\pdfoutput=1
\documentclass[%
 reprint,
superscriptaddress,
nofootinbib,
 amsmath,amssymb,
aps,
pra,
floatfix,
]{revtex4-1}

\usepackage[utf8]{inputenc}
\usepackage[T1]{fontenc}
\usepackage{microtype}

\usepackage{tikz}
\usetikzlibrary{decorations.pathmorphing}
\makeatletter

\def\opticserror#1{\pgfutil@packageerror{tikz/optics}{#1}{}}

\pgfkeys{
  /handlers/.collect unknowns/.style = {
    unknown options/.initial = {},
    .unknown/.code = {%
      \letcs\reserved{pgfk@\pgfkeyscurrentpath/unknown options}%
      \csedef{pgfk@\pgfkeyscurrentpath/unknown options}{%
        \ifx\reserved\empty\else\expandonce\reserved,\fi
        \expandonce\pgfkeyscurrentname
        \ifx\pgfkeysnovalue##1\else=\expandonce\pgfkeyscurrentvalue\fi
      }%
    }
  }
}

\pgfkeys{
  /tikz/.cd,
  optics/.is family,
  optics/.search also={/tikz},
}

\pgfkeys{
  /tikz/use optics/.append code={
  \pgfkeys{
    /tikz/.cd,
      lens/.prefix style={shape=lens,optics,draw},
      slit/.prefix style={shape=slit,optics,draw},
      double slit/.prefix style={shape=double slit,optics,draw},
      thin optics element/.prefix style={shape=thin optics element,optics,draw},
      polarizer/.prefix style={shape=polarizer,optics,draw,object aspect ratio=0.1},
      generic optics io/.prefix style={shape=generic optics io,optics,draw},
      sensor/.prefix style={shape=sensor,optics,object aspect ratio=1,draw},
      sensor line/.prefix style={shape=sensor line,optics,draw},
      mirror/.prefix style={shape=mirror,optics,draw},
      spherical mirror/.prefix style={shape=spherical mirror,optics,draw},
      thick optics element/.prefix style={shape=thick optics element,optics,draw,object aspect ratio=0.1},
      heat filter/.prefix style={shape=thick optics element,optics,draw, object aspect ratio=0.05},
      double amici prism/.prefix style={shape=double amici prism,optics,draw, prism height=1cm, prism apex angle=60},
      screen/.prefix style={optics,/tikz/optics/screen},
      diffraction grating/.style={optics,/tikz/optics/diffraction grating},
      grid/.style={optics, /tikz/optics/grid},
      semi-transparent mirror/.style={optics, /tikz/optics/semi-transparent mirror},
      diaphragm/.style={optics, /tikz/optics/diaphragm},
      beam splitter/.prefix style={optics,/tikz/optics/beam splitter},
      generic lamp/.prefix style={optics,/tikz/optics/generic lamp},
      generic sensor/.prefix style={optics,/tikz/optics/generic sensor},
      halogen lamp/.prefix style={optics,/tikz/optics/halogen lamp},
      spectral lamp/.prefix style={optics,/tikz/optics/spectral lamp},
      laser/.prefix style={optics,/tikz/optics/laser},
      laser'/.prefix style={optics,/tikz/optics/laser'},
      concave mirror/.prefix style={optics, /tikz/optics/concave mirror},
      convex mirror/.prefix style={optics, /tikz/optics/convex mirror},
      ->-/.prefix style={optics,/tikz/optics/->-={##1}},
      -<-/.prefix style={optics,/tikz/optics/-<-={##1}},
      ->>-/.prefix style={optics,/tikz/optics/->>-={##1}},
      -<<-/.prefix style={optics,/tikz/optics/-<<-={##1}},
      ->>>-/.prefix style={optics,/tikz/optics/->>>-={##1}},
      -<<<-/.prefix style={optics,/tikz/optics/-<<<-={##1}},
      ->>>>-/.prefix style={optics,/tikz/optics/->>>>-={##1}},
      -<<<<-/.prefix style={optics,/tikz/optics/-<<<<-={##1}},
      ->n-/.prefix style={optics,/tikz/optics/->n-={##1}},
      -<n-/.prefix style={optics,/tikz/optics/-<n-={##1}},
    }
  }
}

\usetikzlibrary{decorations,decorations.markings, decorations.pathreplacing}
\usepackage{etoolbox}

\pgfkeys{/tikz/optics/.cd,
  object height/.initial=2cm,
}

\pgfdeclareshape{thin optics element}
{
  \savedanchor{\center}{
    \pgfpointorigin
  }
  \anchor{center}{\center}

  \savedmacro\objectHeight{%
    \edef\objectHeight{\pgfkeysvalueof{/tikz/optics/object height}}%
  }

  \savedanchor{\north}{
    \pgf@x=0cm%
    \pgf@y=\objectHeight%
    \pgf@y=0.5\pgf@y%
  }
  \anchor{north}{\north}

  \savedanchor{\south}{
    \pgf@x=0cm%
    \pgf@y=\objectHeight%
    \pgf@y=-0.5\pgf@y%
  }
  \anchor{south}{\south}

  \anchor{east}{\center}
  \anchor{west}{\center}

  \anchorborder{%
    \pgf@xb=\pgf@x
    \pgf@yb=\pgf@y%
    \south%
    \pgf@xa=\pgf@x
    \pgf@ya=\pgf@y%
    \north%
    \advance\pgf@x by-\pgf@xa%
    \advance\pgf@y by-\pgf@ya%
    \pgf@xc=.5\pgf@x
    \pgf@yc=.5\pgf@y%
    \advance\pgf@xa by\pgf@xc
    \advance\pgf@ya by\pgf@yc%
    \edef\pgf@marshal{%
      \noexpand\pgfpointborderrectangle
      {\noexpand\pgfpoint{\the\pgf@xb}{\the\pgf@yb}}
      {\noexpand\pgfpoint{\the\pgf@xc}{\the\pgf@yc}}%
    }%
    \pgf@process{\pgf@marshal}%
    \advance\pgf@x by\pgf@xa%
    \advance\pgf@y by\pgf@ya%
  }

  \backgroundpath
  {
    \north \pgf@xa=\pgf@x \pgf@ya=\pgf@y
    \south \pgf@xb=\pgf@x \pgf@yb=\pgf@y

    \pgfpathmoveto{\pgfpoint{\pgf@xa}{\pgf@ya}}
    \pgfpathlineto{\pgfpoint{\pgf@xb}{\pgf@yb}}
  }
}

\newif\iftikz@optics@lens@converging
\tikz@optics@lens@convergingtrue

\pgfkeys{/tikz/optics/.cd,
  focal length/.initial=1cm,
  lens height/.initial=0.8,
  lens type/.is choice,
  lens type/converging/.code={\tikz@optics@lens@convergingtrue},
  lens type/diverging/.code={\tikz@optics@lens@convergingfalse}
}
\pgfdeclareshape{lens}
{
  \savedanchor{\center}{
    \pgfpointorigin
  }
  \anchor{center}{\center}

  \savedmacro\focalLength{%
    \edef\focalLength{\pgfkeysvalueof{/tikz/optics/focal length}}%
  }

  \savedmacro\objectHeight{%
    \edef\objectHeight{\pgfkeysvalueof{/tikz/optics/object height}}%
  }

  \savedmacro\lensHeight{%
    \pgfmathparse{\pgfkeysvalueof{/tikz/optics/lens height}}%
    \ifpgfmathunitsdeclared%
      \pgfmathsetlengthmacro{\lensHeight}{\pgfkeysvalueof{/tikz/optics/lens height}}%
    \else%
      \pgfmathsetlengthmacro{\lensHeight}{\pgfkeysvalueof{/tikz/optics/lens height}*\pgfkeysvalueof{/tikz/optics/object height}}%
    \fi%
  }

  \savedanchor{\lensnorth}{
    \pgfpointorigin
    \pgf@y=\lensHeight%
    \pgf@y=0.5\pgf@y%
  }
  \anchor{lens north}{\lensnorth}

  \savedanchor{\lenssouth}{
    \pgfpointorigin
    \pgf@y=\lensHeight%
    \pgf@y=-0.5\pgf@y%
  }
  \anchor{lens south}{\lenssouth}

  \savedanchor{\north}{
    \pgf@x=0cm%
    \pgf@y=\objectHeight%
    \pgf@y=0.5\pgf@y%
  }
  \anchor{north}{\north}

  \savedanchor{\south}{
    \pgf@x=0cm%
    \pgf@y=\objectHeight%
    \pgf@y=-0.5\pgf@y%
  }
  \anchor{south}{\south}

  \savedanchor{\eastfocal}{
    \pgf@x=\focalLength%
    \pgf@y=0cm%
  }
  \anchor{east focal point}{\eastfocal}
  \anchor{east focus}{\eastfocal}

  \savedanchor{\westfocal}{
    \pgf@x=-\focalLength%
    \pgf@y=0cm%
  }
  \anchor{west focal point}{\westfocal}
  \anchor{west focus}{\westfocal}

  \anchor{east}{\center}
  \anchor{west}{\center}

  \inheritanchorborder[from=thin optics element]

  \backgroundpath
  {
    \north \pgf@xb=\pgf@x \pgf@yb=\pgf@y
    \south \pgf@xa=\pgf@x \pgf@ya=\pgf@y
    
    \pgfpathmoveto{\pgfpoint{\pgf@xa}{\pgf@ya}}
    \pgfpathlineto{\pgfpoint{\pgf@xb}{\pgf@yb}}

    \iftikz@optics@lens@converging
      \pgfsetarrowsstart{lens arrow}
      \pgfsetarrowsend{lens arrow}
    \else
      \pgfsetarrowsstart{lens arrow reversed}
      \pgfsetarrowsend{lens arrow reversed}
    \fi
  }
}

\pgfkeys{/tikz/optics/.cd,
  slit height/.initial=0.1
}

\pgfdeclareshape{slit}
{
  \savedanchor{\center}{
    \pgfpointorigin
  }
  \anchor{center}{\center}
  \anchor{slit center}{\center} 

  \savedmacro\objectHeight{%
    \edef\objectHeight{\pgfkeysvalueof{/tikz/optics/object height}}%
  }

  \savedmacro\slitHeight{%
    \pgfmathparse{\pgfkeysvalueof{/tikz/optics/slit height}}%
    \ifpgfmathunitsdeclared%
      \pgfmathsetlengthmacro{\slitHeight}{\pgfkeysvalueof{/tikz/optics/slit height}}%
    \else%
      \pgfmathsetlengthmacro{\slitHeight}{\pgfkeysvalueof{/tikz/optics/slit height}*\pgfkeysvalueof{/tikz/optics/object height}}%
    \fi%
  }

  \savedanchor{\north}{
    \pgf@x=0cm%
    \pgf@y=\objectHeight%
    \pgf@y=0.5\pgf@y%
  }
  \anchor{north}{\north}

  \savedanchor{\south}{
    \pgf@x=0cm%
    \pgf@y=\objectHeight%
    \pgf@y=-0.5\pgf@y%
  }
  \anchor{south}{\south}

  \savedanchor{\slitnorth}{
    \pgf@x=0cm%
    \pgf@y=\slitHeight%
    \pgf@y=0.5\pgf@y%
  }
  \anchor{slit north}{\slitnorth}

  \savedanchor{\slitsouth}{
    \pgf@x=0cm%
    \pgf@y=\slitHeight%
    \pgf@y=-0.5\pgf@y%
  }
  \anchor{slit south}{\slitsouth}

  \anchor{east}{\center}
  \anchor{west}{\center}

  \inheritanchorborder[from=thin optics element]

  \backgroundpath
  {
    \pgfmathparse{notless(\slitHeight, \objectHeight)}
    \ifnum\pgfmathresult=1
      \opticserror{<slit height> should be strictly lower than <object height> (in slit)}
    \fi
    \north \pgf@xa=\pgf@x \pgf@ya=\pgf@y
    \slitnorth \pgf@xb=\pgf@x \pgf@yb=\pgf@y
    \pgfpathmoveto{\pgfpoint{\pgf@xa}{\pgf@ya}}
    \pgfpathlineto{\pgfpoint{\pgf@xb}{\pgf@yb}}

    \south \pgf@xa=\pgf@x \pgf@ya=\pgf@y
    \slitsouth \pgf@xb=\pgf@x \pgf@yb=\pgf@y
    \pgfpathmoveto{\pgfpoint{\pgf@xa}{\pgf@ya}}
    \pgfpathlineto{\pgfpoint{\pgf@xb}{\pgf@yb}}
  }
}

\pgfkeys{/tikz/optics/.cd,
  object height/.initial=2cm,
  slit height/.initial=0.075,
  slit separation/.initial=0.2
}

\pgfdeclareshape{double slit}
{
  \savedanchor{\center}{
    \pgfpointorigin
  }
  \anchor{center}{\center}
  
  \savedmacro\objectHeight{%
    \edef\objectHeight{\pgfkeysvalueof{/tikz/optics/object height}}%
  }

  \savedmacro\slitHeight{%
    \pgfmathparse{\pgfkeysvalueof{/tikz/optics/slit height}}%
    \ifpgfmathunitsdeclared%
      \pgfmathsetlengthmacro{\slitHeight}{\pgfkeysvalueof{/tikz/optics/slit height}}%
    \else%
      \pgfmathsetlengthmacro{\slitHeight}{\pgfkeysvalueof{/tikz/optics/slit height}*\pgfkeysvalueof{/tikz/optics/object height}}%
    \fi%
  }

  \savedmacro\slitSeparation{%
    \pgfmathparse{\pgfkeysvalueof{/tikz/optics/slit separation}}%
    \ifpgfmathunitsdeclared%
      \pgfmathsetlengthmacro{\slitSeparation}{\pgfkeysvalueof{/tikz/optics/slit separation}}%
    \else%
      \pgfmathsetlengthmacro{\slitSeparation}{\pgfkeysvalueof{/tikz/optics/slit separation}*\pgfkeysvalueof{/tikz/optics/object height}}%
    \fi%
  }

  \savedmacro\macro@slitOneCenter{
    \def\macro@slitOneCenter{
      \pgfpointorigin
      \pgf@ya=\slitSeparation
      \pgf@ya=0.5\pgf@ya
      \advance \pgf@y by \pgf@ya
    }
  }

  \savedmacro\macro@slitTwoCenter{
    \def\macro@slitTwoCenter{
      \pgfpointorigin
      \pgf@ya=\slitSeparation
      \pgf@ya=-0.5\pgf@ya
      \advance \pgf@y by \pgf@ya
    }
  }

  \savedanchor{\north}{
    \pgf@x=0cm%
    \pgf@y=\objectHeight%
    \pgf@y=0.5\pgf@y%
  }
  \anchor{north}{\north}

  \savedanchor{\south}{
    \pgf@x=0cm%
    \pgf@y=\objectHeight%
    \pgf@y=-0.5\pgf@y%
  }
  \anchor{south}{\south}

  \savedanchor{\slitOneCenter}{
    \macro@slitOneCenter
  }
  \anchor{slit 1 center}{\slitOneCenter}

  \savedanchor{\slitOneNorth}{
    \macro@slitOneCenter
    \pgf@ya = \slitHeight
    \pgf@ya = 0.5\pgf@ya
    \advance \pgf@y by \pgf@ya
  }
  \anchor{slit 1 north}{\slitOneNorth}

  \savedanchor{\slitOneSouth}{
    \macro@slitOneCenter
    \pgf@ya = \slitHeight
    \pgf@ya = -0.5\pgf@ya
    \advance \pgf@y by \pgf@ya
  }
  \anchor{slit 1 south}{\slitOneSouth}

  \savedanchor{\slitTwoCenter}{
    \macro@slitTwoCenter
  }
  \anchor{slit 2 center}{\slitTwoCenter}

  \savedanchor{\slitTwoNorth}{
    \macro@slitTwoCenter
    \pgf@ya = \slitHeight
    \pgf@ya = 0.5\pgf@ya
    \advance \pgf@y by \pgf@ya
  }
  \anchor{slit 2 north}{\slitTwoNorth}

  \savedanchor{\slitTwoSouth}{
    \macro@slitTwoCenter
    \pgf@ya = \slitHeight
    \pgf@ya = -0.5\pgf@ya
    \advance \pgf@y by \pgf@ya
  }
  \anchor{slit 2 south}{\slitTwoSouth}

  \anchor{east}{\center}
  \anchor{west}{\center}

  \inheritanchorborder[from=thin optics element]

  \backgroundpath
  {
    \pgfmathparse{notgreater(\slitSeparation, \slitHeight)}
    \ifnum\pgfmathresult=1
      \opticserror{<slit separation> should be strictly lower than <slit height> (in double slit)}
    \fi
    \pgfmathparse{notless(\slitSeparation+\slitHeight, \objectHeight)}
    \ifnum\pgfmathresult=1
      \opticserror{<slit height> plus <slit separation> should be strictly lower than <object height> (in double slit)}
    \fi
    \north \pgf@xa=\pgf@x \pgf@ya=\pgf@y
    \slitOneNorth \pgf@xb=\pgf@x \pgf@yb=\pgf@y
    \pgfpathmoveto{\pgfpoint{\pgf@xa}{\pgf@ya}}
    \pgfpathlineto{\pgfpoint{\pgf@xb}{\pgf@yb}}

    \slitOneSouth \pgf@xa=\pgf@x \pgf@ya=\pgf@y
    \slitTwoNorth \pgf@xb=\pgf@x \pgf@yb=\pgf@y
    \pgfpathmoveto{\pgfpoint{\pgf@xa}{\pgf@ya}}
    \pgfpathlineto{\pgfpoint{\pgf@xb}{\pgf@yb}}

    \slitTwoSouth \pgf@xa=\pgf@x \pgf@ya=\pgf@y
    \south \pgf@xb=\pgf@x \pgf@yb=\pgf@y
    \pgfpathmoveto{\pgfpoint{\pgf@xa}{\pgf@ya}}
    \pgfpathlineto{\pgfpoint{\pgf@xb}{\pgf@yb}}
  }
}

\pgfkeys{/tikz/optics/.cd,
  object height/.initial=2cm,
  mirror decoration separation/.initial=0.15cm,
  mirror decoration amplitude/.initial=0.125cm,
}

\pgfdeclareshape{mirror}
{
  \savedanchor{\center}{
    \pgfpointorigin
  }
  \anchor{center}{\center}

  \savedmacro\objectHeight{%
    \edef\objectHeight{\pgfkeysvalueof{/tikz/optics/object height}}%
  }

  \savedanchor{\north}{
    \pgf@x=0cm%
    \pgf@y=\objectHeight%
    \pgf@y=0.5\pgf@y%
  }
  \anchor{north}{\north}

  \savedanchor{\south}{
    \pgf@x=0cm%
    \pgf@y=\objectHeight%
    \pgf@y=-0.5\pgf@y%
  }
  \anchor{south}{\south}

  \anchor{east}{\center}
  \anchor{west}{\center}

  \inheritanchorborder[from=thin optics element]

  \backgroundpath
  {
    \north \pgf@xa=\pgf@x \pgf@ya=\pgf@y
    \south \pgf@xb=\pgf@x \pgf@yb=\pgf@y
    \pgfmathparse{\pgfkeysvalueof{/tikz/optics/mirror decoration separation}}
    \ifpgfmathunitsdeclared%
      \pgfmathsetlengthmacro{\pgfdecorationsegmentlength}{\pgfkeysvalueof{/tikz/optics/mirror decoration separation}}%
    \else%
      \pgfmathsetlengthmacro{\pgfdecorationsegmentlength}{\pgfkeysvalueof{/tikz/optics/mirror decoration separation}*\pgfkeysvalueof{/tikz/optics/object height}}%
    \fi%
    \pgfmathsetmacro\initialstep{\pgfdecorationsegmentlength} 
    \pgfmathsetmacro\totallength{\pgfkeysvalueof{/tikz/optics/object height}} 
    \pgfmathsetmacro\newstep{\totallength/floor(\totallength/\initialstep)} 
    \pgfmathsetlengthmacro{\pgfdecorationsegmentlength}{\newstep}
    \pgfmathparse{\pgfkeysvalueof{/tikz/optics/mirror decoration amplitude}}
    \ifpgfmathunitsdeclared%
      \pgfmathsetlengthmacro{\pgfdecorationsegmentamplitude}{-1*\pgfkeysvalueof{/tikz/optics/mirror decoration amplitude}}%
    \else%
      \pgfmathsetlengthmacro{\pgfdecorationsegmentamplitude}{-1*\pgfkeysvalueof{/tikz/optics/mirror decoration amplitude}*\pgfkeysvalueof{/tikz/optics/object height}}%
    \fi%
    \pgfpathmoveto{\pgfpoint{\pgf@xa}{\pgf@ya}}
    \pgfpathlineto{\pgfpoint{\pgf@xb}{\pgf@yb}}
    \pgfdecoratecurrentpath{border} %
    \pgfpathmoveto{\pgfpoint{\pgf@xa}{\pgf@ya}}
    \pgfpathlineto{\pgfpoint{\pgf@xb}{\pgf@yb}}
  }
}

\pgfkeys{/tikz/optics/spherical mirror angle/.initial=150}

\newif\iftikz@optics@sphericalmirror@concave
\tikz@optics@sphericalmirror@concavetrue
\pgfkeys{/tikz/optics/.cd,
  spherical mirror type/.is choice,
  spherical mirror type/concave/.code={\tikz@optics@sphericalmirror@concavetrue},
  spherical mirror type/convex/.code={\tikz@optics@sphericalmirror@concavefalse}
}

\newif\iftikz@optics@sphericalmirror@ltr
\tikz@optics@sphericalmirror@ltrtrue
\pgfkeys{/tikz/optics/.cd,
  spherical mirror orientation/.is choice,
  spherical mirror orientation/ltr/.code={\tikz@optics@sphericalmirror@ltrtrue},
  spherical mirror orientation/rtl/.code={\tikz@optics@sphericalmirror@ltrfalse}
}

\pgfkeys{/tikz/optics/.cd,
      concave mirror/.style={optics, spherical mirror, /tikz/optics/spherical mirror type=concave},
      convex mirror/.style={optics, spherical mirror, /tikz/optics/spherical mirror type=convex},
}

\pgfdeclareshape{spherical mirror}{%
  \savedmacro\installsphericalmirrorparameters{%
    \pgfextract@process\centerpoint{%
      \pgfpointorigin
    }%
    \pgfmathsetlengthmacro\height{\pgfkeysvalueof{/tikz/optics/object height}}%
    \pgfmathdeclarefunction{from_radius}{1}{
      \begingroup
      \pgfmathparse{notless(2*#1,\height)}
      \ifnum\pgfmathresult=0
        \opticserror{(in /tikz/optics/spherical mirror angle=from_radius(R)) : for a spherical mirror, the radius R cannot be smaller than half the height </tikz/optics/object height>. Set a bigger radius of a smaller height.}
      \fi
      \newdimen\angle
      \pgfmathsetlength\angle{2*asin(\height/(2*#1))}
      \pgf@x=\angle
      \pgfmathreturn\pgf@x
      \endgroup
    }
    \pgfmathsetmacro\angle{\pgfkeysvalueof{/tikz/optics/spherical mirror angle}}
    \pgfmathsetlengthmacro\radius{\height/(2*sin(\angle/2))}
    \pgfmathmod{\angle}{360}%
    \ifdim\pgfmathresult pt<0pt\relax%
      \pgfmathadd@{\pgfmathresult}{360}%
    \fi%
    \let\angle\pgfmathresult%
    \pgfmathdivide@{\pgfmathresult}{2}%
    \let\halfangle\pgfmathresult%
    \iftikz@optics@sphericalmirror@concave
      \iftikz@optics@sphericalmirror@ltr
        \pgfmathsetmacro\startangle{-\halfangle}
        \pgfmathsetmacro\endangle{+\halfangle}
      \else
        \pgfmathsetmacro\startangle{180-\halfangle}
        \pgfmathsetmacro\endangle{180+\halfangle}
      \fi
      \else
      \iftikz@optics@sphericalmirror@ltr
        \pgfmathsetmacro\startangle{180-\halfangle}
        \pgfmathsetmacro\endangle{180+\halfangle}
      \else
        \pgfmathsetmacro\startangle{-\halfangle}
        \pgfmathsetmacro\endangle{+\halfangle}
      \fi
    \fi
    \pgfmathabs@{\halfangle}%
    \pgfmathcos@{\pgfmathresult}%
    \let\coshalfangle\pgfmathresult%
    \pgfmathabs@{\halfangle}%
    \pgfmathsin@{\pgfmathresult}%
    \let\sinhalfangle\pgfmathresult%
    \pgfmathsetlength\pgf@xa{\radius*\coshalfangle}
    \edef\rcoshalfangle{\the\pgf@xa}%
    \pgfmathsetlength\pgf@xa{\radius*\sinhalfangle}
    \edef\rsinhalfangle{\the\pgf@xa}%
    \pgfextract@process\arcstart{%
      \pgfqpointpolar{\startangle}{\radius}%
      \pgf@xa\pgf@x%
      \pgf@ya\pgf@y%
      \centerpoint%
      \advance\pgf@x\pgf@xa%
      \advance\pgf@y\pgf@ya%
    }%
    \pgfextract@process\arcend{%
      \pgfqpointpolar{\endangle}{\radius}%
      \pgf@xa\pgf@x%
      \pgf@ya\pgf@y%
      \centerpoint%
      \advance\pgf@x\pgf@xa%
      \advance\pgf@y\pgf@ya%
    }%
    \def\convexrtlsetx#1#2{
      \iftikz@optics@sphericalmirror@concave%
        \iftikz@optics@sphericalmirror@ltr%
          \advance\pgf@x by #1
        \else
          \advance\pgf@x by #2
        \fi%
      \else
        \iftikz@optics@sphericalmirror@ltr%
          \advance\pgf@x by #2
        \else
          \advance\pgf@x by #1
        \fi%
      \fi%
    }
    \def\convexrtlinvert{%
      \iftikz@optics@sphericalmirror@concave%
        \iftikz@optics@sphericalmirror@ltr%
        \else
          \pgf@x=-\pgf@x%
        \fi%
      \else
        \iftikz@optics@sphericalmirror@ltr%
          \pgf@x=-\pgf@x%
        \else
        \fi%
      \fi%
    }%
    \addtosavedmacro{\radius}%
    \addtosavedmacro{\rcoshalfangle}%
    \addtosavedmacro{\rsinhalfangle}%
    \addtosavedmacro{\endangle}%
    \addtosavedmacro{\startangle}%
    \addtosavedmacro{\centerpoint}%
    \addtosavedmacro{\arcstart}%
    \addtosavedmacro{\arcend}%
     \addtosavedmacro{\convexrtlinvert}%
  }%
  \savedanchor\mirrorcenterpoint{%
    \pgfpointorigin
  }%
  \savedanchor\centerpoint{%
    \pgfpointorigin%
    \advance\pgf@x by \radius%
    \advance\pgf@x by \rcoshalfangle%
    \iftikz@optics@sphericalmirror@concave%
      \iftikz@optics@sphericalmirror@ltr%
      \else
        \pgf@x=-\pgf@x%
      \fi%
    \else
      \iftikz@optics@sphericalmirror@ltr%
        \pgf@x=-\pgf@x%
      \else
      \fi%
    \fi%
    \divide\pgf@x by 2%
  }%
  \savedanchor\focalpoint{%
    \pgfpointorigin%
    \pgf@xa=\radius%
    \advance\pgf@x by .5\pgf@xa%
    \iftikz@optics@sphericalmirror@concave%
      \iftikz@optics@sphericalmirror@ltr%
      \else
        \pgf@x=-\pgf@x%
      \fi%
    \else
      \iftikz@optics@sphericalmirror@ltr%
        \pgf@x=-\pgf@x%
      \else
      \fi%
    \fi%
  }%
  \savedanchor\north{%
    \pgfpointorigin
    \pgf@xa=0pt
    \advance\pgf@xa by \rcoshalfangle
    \advance\pgf@xa by \radius
    \divide\pgf@xa by 2
    \advance\pgf@x by \pgf@xa
    \advance\pgf@y by \rsinhalfangle
    \convexrtlinvert
  }%
  \savedanchor\arccenter{%
    \centerpoint
    \advance\pgf@x by \radius
    \convexrtlinvert
  }
  \savedanchor\south{%
    \pgfpointorigin
    \pgf@xa=0pt
    \advance\pgf@xa by \rcoshalfangle
    \advance\pgf@xa by \radius
    \divide\pgf@xa by 2
    \advance\pgf@x by \pgf@xa
    \advance\pgf@y by -\rsinhalfangle
    \convexrtlinvert
  }
  \savedanchor\east{%
    \pgfpointorigin
    \convexrtlsetx{\radius}{\rcoshalfangle}
    \convexrtlinvert
  }
  \savedanchor\west{%
    \pgfpointorigin
    \convexrtlsetx{\rcoshalfangle}{\radius}
    \convexrtlinvert
  }
  \savedanchor\northwest{%
    \pgfpointorigin
    \convexrtlsetx{\rcoshalfangle}{\radius}
    \advance\pgf@y by \rsinhalfangle
    \convexrtlinvert
  }
  \savedanchor\southwest{%
    \pgfpointorigin
    \convexrtlsetx{\rcoshalfangle}{\radius}
    \advance\pgf@y by -\rsinhalfangle
    \convexrtlinvert
  }
  \savedanchor\northeast{%
    \pgfpointorigin
    \convexrtlsetx{\radius}{\rcoshalfangle}
    \advance\pgf@y by \rsinhalfangle
    \convexrtlinvert
  }
  \savedanchor\southeast{%
    \pgfpointorigin
    \convexrtlsetx{\radius}{\rcoshalfangle}
    \advance\pgf@y by -\rsinhalfangle
    \convexrtlinvert
  }
  \anchor{arc start}{%
    \installsphericalmirrorparameters%
    \arcstart%
  }
  \anchor{arc end}{%
    \installsphericalmirrorparameters%
    \arcend%
  }
  \anchor{focal point}{%
    \installsphericalmirrorparameters%
    \focalpoint
  }
  \anchor{focus}{%
    \installsphericalmirrorparameters%
    \focalpoint
  }
  \anchor{mirror center}{\mirrorcenterpoint}  
  \anchor{center}{\centerpoint} 
  \anchor{arc center}{\arccenter}
  \anchor{north}{\north}%
  \anchor{south}{\south}%
  \anchor{east}{\east}%
  \anchor{west}{\west}%
  \anchor{north west}{\northwest}%
  \anchor{south west}{\southwest}%
  \anchor{north east}{\northeast}%
  \anchor{south east}{\southeast}%
  \backgroundpath{%
    \installsphericalmirrorparameters%
    \iftikz@optics@sphericalmirror@concave
    \else
    \fi
    \pgfmathparse{\pgfkeysvalueof{/tikz/optics/mirror decoration separation}}
    \ifpgfmathunitsdeclared%
      \pgfmathsetlengthmacro{\pgfdecorationsegmentlength}{\pgfkeysvalueof{/tikz/optics/mirror decoration separation}}%
    \else%
      \pgfmathsetlengthmacro{\pgfdecorationsegmentlength}{\pgfkeysvalueof{/tikz/optics/mirror decoration separation}*\pgfkeysvalueof{/tikz/optics/object height}}%
    \fi%
    \pgfmathsetmacro\initialstep{\pgfdecorationsegmentlength} 
    \pgfmathsetmacro\totallength{(2*pi/360)*\angle*\radius} 
    \pgfmathsetmacro\newstep{\totallength/floor(\totallength/\initialstep)} 
    \pgfmathsetlengthmacro{\pgfdecorationsegmentlength}{\newstep}
    \pgfmathparse{\pgfkeysvalueof{/tikz/optics/mirror decoration amplitude}}
    \ifpgfmathunitsdeclared%
      \pgfmathsetlengthmacro{\pgfdecorationsegmentamplitude}{-1*\pgfkeysvalueof{/tikz/optics/mirror decoration amplitude}}%
    \else%
      \pgfmathsetlengthmacro{\pgfdecorationsegmentamplitude}{-1*\pgfkeysvalueof{/tikz/optics/mirror decoration amplitude}*\pgfkeysvalueof{/tikz/optics/object height}}%
    \fi%
    \pgfpathmoveto{\arcstart}%
    \pgfpatharc{\startangle}{\endangle}{\radius}%
    \pgfdecoratecurrentpath{border} %
    \pgfpathmoveto{\arcstart}%
    \pgfpatharc{\startangle}{\endangle}{\radius}%
  }
  \anchorborder{%
    \edef\externalx{\the\pgf@x}%
    \edef\externaly{\the\pgf@y}%
    \installsphericalmirrorparameters%
    \pgfpointborderellipse{ \pgfpoint{\externalx}{\externaly} }{ \pgfpoint{\radius}{\radius} }%
  }%
}

\pgfkeys{/tikz/optics/.cd,
  object aspect ratio/.initial=0.2,
  object width/.style={/tikz/optics/object aspect ratio=#1}
}

\pgfdeclareshape{thick optics element}
{
  \savedanchor{\center}{
    \pgfpointorigin
  }
  \anchor{center}{\center}

  \savedmacro\objectHeight{%
    \edef\objectHeight{\pgfkeysvalueof{/tikz/optics/object height}}%
  }

  \savedmacro\objectWidth{%
    \pgfmathparse{\pgfkeysvalueof{/tikz/optics/object aspect ratio}}
    \ifpgfmathunitsdeclared%
      \pgfmathsetlengthmacro{\objectWidth}{\pgfkeysvalueof{/tikz/optics/object aspect ratio}}%
    \else%
      \pgfmathsetlengthmacro{\objectWidth}{\pgfkeysvalueof{/tikz/optics/object aspect ratio}*\pgfkeysvalueof{/tikz/optics/object height}}%
    \fi%
  }

  \savedanchor{\northeast}{
    \pgf@x=\objectWidth%
    \pgf@y=\objectHeight%
    \pgf@y=0.5\pgf@y%
    \pgf@x=0.5\pgf@x%
  }
  \anchor{north east}{\northeast}

  \savedanchor{\southwest}{
    \pgf@x=\objectWidth%
    \pgf@y=\objectHeight%
    \pgf@y=-0.5\pgf@y%
    \pgf@x=-0.5\pgf@x%
  }
  \anchor{south west}{\southwest}

  \anchor{north}{
    \center \pgf@xa=\pgf@x \pgf@ya=\pgf@y
    \northeast \pgf@xb=\pgf@x \pgf@yb=\pgf@y
    \pgf@x=\pgf@xa
    \pgf@y=\pgf@yb
  }

  \anchor{south}{
    \center \pgf@xa=\pgf@x \pgf@ya=\pgf@y
    \southwest \pgf@xb=\pgf@x \pgf@yb=\pgf@y
    \pgf@x=\pgf@xa
    \pgf@y=\pgf@yb
  }

  \anchor{east}{
    \center \pgf@xa=\pgf@x \pgf@ya=\pgf@y
    \northeast \pgf@xb=\pgf@x \pgf@yb=\pgf@y
    \pgf@x=\pgf@xb
    \pgf@y=\pgf@ya
  }

  \anchor{west}{
    \center \pgf@xa=\pgf@x \pgf@ya=\pgf@y
    \southwest \pgf@xb=\pgf@x \pgf@yb=\pgf@y
    \pgf@x=\pgf@xb
    \pgf@y=\pgf@ya
  }

  \anchor{north west}{
    \northeast \pgf@xa=\pgf@x \pgf@ya=\pgf@y
    \southwest \pgf@xb=\pgf@x \pgf@yb=\pgf@y
    \pgf@x=\pgf@xb
    \pgf@y=\pgf@ya
  }

  \anchor{south east}{
    \northeast \pgf@xa=\pgf@x \pgf@ya=\pgf@y
    \southwest \pgf@xb=\pgf@x \pgf@yb=\pgf@y
    \pgf@x=\pgf@xa
    \pgf@y=\pgf@yb
  }

  \inheritanchorborder[from=rectangle]

  \backgroundpath
  {
    \pgfpathrectanglecorners{\northeast}{\southwest}
  }
}

\pgfdeclareshape{polarizer}
{
  \savedanchor{\center}{
    \pgfpointorigin
  }
  \anchor{center}{\center}

  \savedmacro\objectHeight{%
    \edef\objectHeight{\pgfkeysvalueof{/tikz/optics/object height}}%
  }

  \savedmacro\objectWidth{%
    \pgfmathparse{\pgfkeysvalueof{/tikz/optics/object aspect ratio}}
    \ifpgfmathunitsdeclared%
      \pgfmathsetlengthmacro{\objectWidth}{\pgfkeysvalueof{/tikz/optics/object aspect ratio}}%
    \else%
      \pgfmathsetlengthmacro{\objectWidth}{\pgfkeysvalueof{/tikz/optics/object aspect ratio}*\pgfkeysvalueof{/tikz/optics/object height}}%
    \fi%
  }

  \savedanchor{\northeast}{
    \pgf@x=\objectWidth%
    \pgf@y=\objectHeight%
    \pgf@y=0.5\pgf@y%
    \pgf@x=0.5\pgf@x%
  }
  \anchor{north east}{\northeast}

  \savedanchor{\southwest}{
    \pgf@x=\objectWidth%
    \pgf@y=\objectHeight%
    \pgf@y=-0.5\pgf@y%
    \pgf@x=-0.5\pgf@x%
  }
  \anchor{south west}{\southwest}

  \anchor{north}{
    \center \pgf@xa=\pgf@x \pgf@ya=\pgf@y
    \northeast \pgf@xb=\pgf@x \pgf@yb=\pgf@y
    \pgf@x=\pgf@xa
    \pgf@y=\pgf@yb
  }

  \anchor{south}{
    \center \pgf@xa=\pgf@x \pgf@ya=\pgf@y
    \southwest \pgf@xb=\pgf@x \pgf@yb=\pgf@y
    \pgf@x=\pgf@xa
    \pgf@y=\pgf@yb
  }

  \anchor{east}{
    \center \pgf@xa=\pgf@x \pgf@ya=\pgf@y
    \northeast \pgf@xb=\pgf@x \pgf@yb=\pgf@y
    \pgf@x=\pgf@xb
    \pgf@y=\pgf@ya
  }

  \anchor{west}{
    \center \pgf@xa=\pgf@x \pgf@ya=\pgf@y
    \southwest \pgf@xb=\pgf@x \pgf@yb=\pgf@y
    \pgf@x=\pgf@xb
    \pgf@y=\pgf@ya
  }

  \anchor{north west}{
    \northeast \pgf@xa=\pgf@x \pgf@ya=\pgf@y
    \southwest \pgf@xb=\pgf@x \pgf@yb=\pgf@y
    \pgf@x=\pgf@xb
    \pgf@y=\pgf@ya
  }

  \anchor{south east}{
    \northeast \pgf@xa=\pgf@x \pgf@ya=\pgf@y
    \southwest \pgf@xb=\pgf@x \pgf@yb=\pgf@y
    \pgf@x=\pgf@xa
    \pgf@y=\pgf@yb
  }

  \inheritanchorborder[from=rectangle]

  \backgroundpath
  {
    \pgfpathrectanglecorners{\northeast}{\southwest}

    \northeast \pgf@xa=\pgf@x \pgf@ya=\pgf@y
    \southwest \pgf@xb=\pgf@x \pgf@yb=\pgf@y
    \pgfpathmoveto{\pgfpoint{\pgf@xa}{\pgf@ya}}
    \pgfpathlineto{\pgfpoint{\pgf@xb}{\pgf@yb}}
  }
}

\pgfkeys{/tikz/optics/.cd,
  prism height/.initial=1.5cm,
  prism apex angle/.initial=60,
}
\pgfdeclareshape{double amici prism}
{
  \savedanchor{\center}{
    \pgfpointorigin
  }
  \anchor{center}{\center}

  \savedmacro\prismHeight{%
    \edef\prismHeight{\pgfkeysvalueof{/tikz/optics/prism height}}%
  }

  \savedmacro\apexAngle{%
      \pgfmathsetlengthmacro{\apexAngle}{\pgfkeysvalueof{/tikz/optics/prism apex angle}}
  }

  \savedmacro\demiPrismWidth{%
      \pgfmathsetlengthmacro{\demiPrismWidth}{tan(0.5*\pgfkeysvalueof{/tikz/optics/prism apex angle})*\pgfkeysvalueof{/tikz/optics/prism height}}
  }

  \savedanchor{\northeast}{
    \pgf@x=\demiPrismWidth%
    \pgf@y=\prismHeight%
    \pgf@y=0.5\pgf@y%
  }
  \anchor{north east}{\northeast}

  \savedanchor{\southwest}{
    \pgf@x=\demiPrismWidth%
    \pgf@y=\prismHeight%
    \pgf@x=-2\pgf@x%
    \pgf@y=-0.5\pgf@y%
  }
  \anchor{south west}{\southwest}

  \savedanchor{\northwest}{
    \pgf@x=\demiPrismWidth%
    \pgf@y=\prismHeight%
    \pgf@x=-\pgf@x%
    \pgf@y=0.5\pgf@y%
  }
  \anchor{north west}{\northwest}

  \savedanchor{\southeast}{
    \pgf@x=\demiPrismWidth%
    \pgf@y=\prismHeight%
    \pgf@x=2\pgf@x%
    \pgf@y=-0.5\pgf@y%
  }
  \anchor{south east}{\southeast}

  \anchor{north}{
    \center \pgf@xa=\pgf@x \pgf@ya=\pgf@y
    \northeast \pgf@xb=\pgf@x \pgf@yb=\pgf@y
    \pgf@x=\pgf@xa
    \pgf@y=\pgf@yb
  }

  \savedanchor{\south}{
    \pgfpointorigin
    \pgf@y=\prismHeight%
    \pgf@y=-0.5\pgf@y%
  }
  \anchor{south}{\south}

  \anchor{east}{
    \southeast \pgf@xa=\pgf@x \pgf@ya=\pgf@y
    \northeast \pgf@xb=\pgf@x \pgf@yb=\pgf@y
    \pgf@x=\pgf@xa
    \advance\pgf@x by\pgf@xb
    \pgf@x=0.5\pgf@x
    \pgf@y=\pgf@ya
    \advance\pgf@y by\pgf@yb
    \pgf@y=0.5\pgf@y
  }

  \anchor{west}{
    \southwest \pgf@xa=\pgf@x \pgf@ya=\pgf@y
    \northwest \pgf@xb=\pgf@x \pgf@yb=\pgf@y
    \pgf@x=\pgf@xa
    \advance\pgf@x by\pgf@xb
    \pgf@x=0.5\pgf@x
    \pgf@y=\pgf@ya
    \advance\pgf@y by\pgf@yb
    \pgf@y=0.5\pgf@y
  }

   \inheritanchorborder[from=rectangle]

  \backgroundpath
  {
    \northwest
    \pgfpathmoveto{\pgfpoint{\pgf@x}{\pgf@y}}
    \southwest
    \pgfpathlineto{\pgfpoint{\pgf@x}{\pgf@y}}
    \southeast
    \pgfpathlineto{\pgfpoint{\pgf@x}{\pgf@y}}
    \northeast
    \pgfpathlineto{\pgfpoint{\pgf@x}{\pgf@y}}
    \pgfpathclose
    \pgfsetbeveljoin
    \northwest
    \pgfpathmoveto{\pgfpoint{\pgf@x}{\pgf@y}}
    \south
    \pgfpathlineto{\pgfpoint{\pgf@x}{\pgf@y}}
    \northeast
    \pgfpathlineto{\pgfpoint{\pgf@x}{\pgf@y}}
  }
}

\pgfkeys{/tikz/optics/.cd,
  io body height/.initial=0.75cm,
  io body aspect ratio/.initial=2,
  io aperture width/.initial=0.33,
  io aperture height/.initial=0.66,
  io aperture shift/.initial=0,
}
\pgfkeys{/tikz/optics/io body width/.style={/tikz/optics/io body aspect ratio=#1}}
\newif\if@tikz@optics@io@ltr
\@tikz@optics@io@ltrtrue
\pgfkeys{/tikz/optics/io orientation/.is choice}
\pgfkeys{/tikz/optics/io orientation/ltr/.code={\@tikz@optics@io@ltrtrue}}
\pgfkeys{/tikz/optics/io orientation/rtl/.code={\@tikz@optics@io@ltrfalse}}
\pgfdeclareshape{generic optics io}
{
  \savedanchor{\center}{
    \pgfpointorigin
  }
  \anchor{center}{\center}

  \savedmacro\objectHeight{%
    \edef\objectHeight{\pgfkeysvalueof{/tikz/optics/io body height}}%
  }

  \savedmacro\objectWidth{%
    \pgfmathparse{\pgfkeysvalueof{/tikz/optics/io body aspect ratio}}
    \ifpgfmathunitsdeclared%
      \pgfmathsetlengthmacro{\objectWidth}{\pgfkeysvalueof{/tikz/optics/io body aspect ratio}}%
    \else%
      \pgfmathsetlengthmacro{\objectWidth}{\pgfkeysvalueof{/tikz/optics/io body aspect ratio}*\pgfkeysvalueof{/tikz/optics/io body height}}%
    \fi%
  }

  \savedmacro\outHeight{%
    \pgfmathparse{\pgfkeysvalueof{/tikz/optics/io aperture height}}
    \ifpgfmathunitsdeclared%
      \pgfmathsetlengthmacro{\outHeight}{\pgfkeysvalueof{/tikz/optics/io aperture height}}%
    \else%
      \pgfmathsetlengthmacro{\outHeight}{\pgfkeysvalueof{/tikz/optics/io aperture height}*\pgfkeysvalueof{/tikz/optics/io body height}}%
    \fi%
  }

  \savedmacro\outWidth{%
    \pgfmathparse{\pgfkeysvalueof{/tikz/optics/io aperture width}}
    \ifpgfmathunitsdeclared%
      \pgfmathsetlengthmacro{\outWidth}{\pgfkeysvalueof{/tikz/optics/io aperture width}}%
    \else%
      \pgfmathsetlengthmacro{\outWidth}{\pgfkeysvalueof{/tikz/optics/io aperture width}*\pgfkeysvalueof{/tikz/optics/io body height}}%
    \fi%
  }

  \savedmacro\outShift{%
    \pgfmathparse{\pgfkeysvalueof{/tikz/optics/io aperture shift}}
    \ifpgfmathunitsdeclared%
      \pgfmathsetlengthmacro{\outShift}{\pgfkeysvalueof{/tikz/optics/io aperture shift}}%
    \else%
      \pgfmathsetlengthmacro{\outShift}{\pgfkeysvalueof{/tikz/optics/io aperture shift}*\pgfkeysvalueof{/tikz/optics/io body height}}%
    \fi%
  }

  \savedanchor{\bodynortheast}{
    \pgf@x=\objectWidth%
    \pgf@y=\objectHeight%
    \pgf@y=0.5\pgf@y%
    \pgf@x=0.5\pgf@x%
  }
  \anchor{body north east}{\bodynortheast}

  \savedanchor{\bodysouthwest}{
    \pgf@x=\objectWidth%
    \pgf@y=\objectHeight%
    \pgf@y=-0.5\pgf@y%
    \pgf@x=-0.5\pgf@x%
  }
  \anchor{body south west}{\bodysouthwest}

  \savedanchor{\outnortheast}{   
    \if@tikz@optics@io@ltr
      \pgf@x=\objectWidth%
      \pgf@x=0.5\pgf@x%
      \advance\pgf@x by\outWidth
    \else
      \pgf@x=\objectWidth%
      \pgf@x=0.5\pgf@x%
      \pgf@x=-\pgf@x%
    \fi
    \pgf@y=0pt%
    \advance\pgf@y by\outHeight
    \pgf@y=0.5\pgf@y%
    \advance\pgf@y by\outShift
  }
  \anchor{aperture north east}{\outnortheast}

  \savedanchor{\outsouthwest}{
    \if@tikz@optics@io@ltr
      \pgf@x=\objectWidth%
      \pgf@x=0.5\pgf@x%
    \else
      \pgf@x=\objectWidth%
      \pgf@x=0.5\pgf@x%
      \advance\pgf@x by\outWidth
      \pgf@x=-\pgf@x%
    \fi
    \pgf@y=0pt%
    \advance\pgf@y by\outHeight
    \pgf@y=-0.5\pgf@y%
    \advance\pgf@y by\outShift
  }
  \anchor{aperture south west}{\outsouthwest}

  \savedanchor{\outcenter}{
    \if@tikz@optics@io@ltr
      \pgf@x=\objectWidth%
      \advance\pgf@x by\outWidth
    \else
      \pgf@x=\objectWidth%
      \advance\pgf@x by\outWidth
      \pgf@x=-\pgf@x%
    \fi
    \pgf@x=0.5\pgf@x%
    \pgf@y=0pt%
    \advance\pgf@y by\outShift
  }
  \anchor{aperture center}{\outcenter}

  \anchor{body center}{\center}

  \anchor{text}{
    \pgfpointorigin
    \advance\pgf@x by -.5\wd\pgfnodeparttextbox%
    \advance\pgf@y by -.5\ht\pgfnodeparttextbox%
    \advance\pgf@y by +.5\dp\pgfnodeparttextbox%
  }

  \anchor{body north}{
    \center \pgf@xa=\pgf@x \pgf@ya=\pgf@y
    \bodynortheast \pgf@xb=\pgf@x \pgf@yb=\pgf@y
    \pgf@x=\pgf@xa
    \pgf@y=\pgf@yb
  }
  \anchor{north}{
    \center \pgf@xa=\pgf@x \pgf@ya=\pgf@y
    \bodynortheast \pgf@xb=\pgf@x \pgf@yb=\pgf@y
    \pgf@x=\pgf@xa
    \pgf@y=\pgf@yb
  }

  \anchor{body south}{
    \center \pgf@xa=\pgf@x \pgf@ya=\pgf@y
    \bodysouthwest \pgf@xb=\pgf@x \pgf@yb=\pgf@y
    \pgf@x=\pgf@xa
    \pgf@y=\pgf@yb
  }
  \anchor{south}{
    \center \pgf@xa=\pgf@x \pgf@ya=\pgf@y
    \bodysouthwest \pgf@xb=\pgf@x \pgf@yb=\pgf@y
    \pgf@x=\pgf@xa
    \pgf@y=\pgf@yb
  }

  \anchor{body east}{
    \center \pgf@xa=\pgf@x \pgf@ya=\pgf@y
    \bodynortheast \pgf@xb=\pgf@x \pgf@yb=\pgf@y
    \pgf@x=\pgf@xb
    \pgf@y=\pgf@ya
  }

  \anchor{body west}{
    \center \pgf@xa=\pgf@x \pgf@ya=\pgf@y
    \bodysouthwest \pgf@xb=\pgf@x \pgf@yb=\pgf@y
    \pgf@x=\pgf@xb
    \pgf@y=\pgf@ya
  }

  \anchor{body north west}{
    \bodynortheast \pgf@xa=\pgf@x \pgf@ya=\pgf@y
    \bodysouthwest \pgf@xb=\pgf@x \pgf@yb=\pgf@y
    \pgf@x=\pgf@xb
    \pgf@y=\pgf@ya
  }

  \anchor{body south east}{
    \bodynortheast \pgf@xa=\pgf@x \pgf@ya=\pgf@y
    \bodysouthwest \pgf@xb=\pgf@x \pgf@yb=\pgf@y
    \pgf@x=\pgf@xa
    \pgf@y=\pgf@yb
  }

  \anchor{aperture north}{
    \outcenter \pgf@xa=\pgf@x \pgf@ya=\pgf@y
    \outnortheast \pgf@xb=\pgf@x \pgf@yb=\pgf@y
    \pgf@x=\pgf@xa
    \pgf@y=\pgf@yb
  }

  \anchor{aperture south}{
    \outcenter \pgf@xa=\pgf@x \pgf@ya=\pgf@y
    \outsouthwest \pgf@xb=\pgf@x \pgf@yb=\pgf@y
    \pgf@x=\pgf@xa
    \pgf@y=\pgf@yb
  }

  \anchor{aperture east}{
    \outcenter \pgf@xa=\pgf@x \pgf@ya=\pgf@y
    \outnortheast \pgf@xb=\pgf@x \pgf@yb=\pgf@y
    \pgf@x=\pgf@xb
    \pgf@y=\pgf@ya
  }

  \anchor{aperture west}{
    \outcenter \pgf@xa=\pgf@x \pgf@ya=\pgf@y
    \outsouthwest \pgf@xb=\pgf@x \pgf@yb=\pgf@y
    \pgf@x=\pgf@xb
    \pgf@y=\pgf@ya
  }

  \anchor{aperture north west}{
    \outnortheast \pgf@xa=\pgf@x \pgf@ya=\pgf@y
    \outsouthwest \pgf@xb=\pgf@x \pgf@yb=\pgf@y
    \pgf@x=\pgf@xb
    \pgf@y=\pgf@ya
  }

  \anchor{aperture south east}{
    \outnortheast \pgf@xa=\pgf@x \pgf@ya=\pgf@y
    \outsouthwest \pgf@xb=\pgf@x \pgf@yb=\pgf@y
    \pgf@x=\pgf@xa
    \pgf@y=\pgf@yb
  }

  \anchor{center}{\center}

  \savedanchor{\realeast}{
    \pgfpointorigin
    \if@tikz@optics@io@ltr
      \pgf@x=\objectWidth%
      \pgf@x=0.5\pgf@x%
      \advance\pgf@x by\outWidth
    \else
      \pgf@x=\objectWidth%
      \pgf@x=0.5\pgf@x%
    \fi
  }
  \anchor{east}{\realeast}

  \savedanchor{\realwest}{
    \pgfpointorigin
    \if@tikz@optics@io@ltr
      \pgf@x=\objectWidth%
      \pgf@x=-0.5\pgf@x%
    \else
      \pgf@x=\objectWidth%
      \pgf@x=0.5\pgf@x%
      \advance\pgf@x by\outWidth
      \pgf@x=-\pgf@x%
    \fi
  }
  \anchor{west}{\realwest}

  \savedanchor{\anchorbordersouthwest}{
    \if@tikz@optics@io@ltr
      \pgf@x=\objectWidth%
      \pgf@x=-0.5\pgf@x%
    \else
      \pgf@x=\objectWidth%
      \pgf@x=0.5\pgf@x%
      \advance\pgf@x by\outWidth
      \pgf@x=-\pgf@x%
    \fi
    \pgf@y=\objectHeight%
    \pgf@y=-0.5\pgf@y%
  }

  \savedanchor{\anchorbordernortheast}{
    \if@tikz@optics@io@ltr
      \pgf@x=\objectWidth%
      \pgf@x=0.5\pgf@x%
      \advance\pgf@x by\outWidth
    \else
      \pgf@x=\objectWidth%
      \pgf@x=0.5\pgf@x%
    \fi
    \pgf@y=\objectHeight%
    \pgf@y=0.5\pgf@y%
  }

  \anchorborder{%
    \pgf@xb=\pgf@x
    \pgf@yb=\pgf@y%
    \anchorbordersouthwest%
    \pgf@xa=\pgf@x
    \pgf@ya=\pgf@y%
    \anchorbordernortheast%
    \advance\pgf@x by-\pgf@xa%
    \advance\pgf@y by-\pgf@ya%
    \pgf@xc=.5\pgf@x
    \pgf@yc=.5\pgf@y%
    \advance\pgf@xa by\pgf@xc
    \advance\pgf@ya by\pgf@yc%
    \edef\pgf@marshal{%
      \noexpand\pgfpointborderrectangle
      {\noexpand\pgfqpoint{\the\pgf@xb}{\the\pgf@yb}}
      {\noexpand\pgfqpoint{\the\pgf@xc}{\the\pgf@yc}}%
    }%
    \pgf@process{\pgf@marshal}%
  }

  \backgroundpath
  {
    \pgfpathrectanglecorners{\bodynortheast}{\bodysouthwest}
    \pgfpathrectanglecorners{\outnortheast}{\outsouthwest}

  }
}

\pgfkeys{/tikz/optics/.cd,
  sensor line height/.initial=2cm,
  sensor line aspect ratio/.initial=0.2,
  sensor line pixel number/.initial=5,
  sensor line pixel width/.initial=0.4,
  sensor line inner ysep/.initial=0.05, 
}

\pgfdeclareshape{sensor line}
{
  \savedanchor{\center}{
    \pgfpointorigin
  }
  \anchor{center}{\center}

  \savedmacro\objectHeight{%
    \edef\objectHeight{\pgfkeysvalueof{/tikz/optics/sensor line height}}%
  }

  \savedmacro\objectWidth{%
    \pgfmathsetlengthmacro{\objectWidth}{\pgfkeysvalueof{/tikz/optics/sensor line aspect ratio}*\pgfkeysvalueof{/tikz/optics/sensor line height}}
  }

  \savedmacro\pixelNumber{%
    \pgfmathtruncatemacro\pixelNumber{\pgfkeysvalueof{/tikz/optics/sensor line pixel number}}%
  }

  \savedmacro\innerysep{%
    \pgfmathparse{\pgfkeysvalueof{/tikz/optics/sensor line inner ysep}}
    \ifpgfmathunitsdeclared%
      \pgfmathsetlengthmacro{\innerysep}{\pgfkeysvalueof{/tikz/optics/sensor line inner ysep}}%
    \else%
      \pgfmathsetlengthmacro{\innerysep}{\pgfkeysvalueof{/tikz/optics/sensor line inner ysep}*\pgfkeysvalueof{/tikz/optics/sensor line height}}%
    \fi%
  }
  
  \savedmacro\pixelWidth{%
    \pgfmathparse{\pgfkeysvalueof{/tikz/optics/sensor line pixel width}}
    \ifpgfmathunitsdeclared%
      \pgfmathsetlengthmacro{\pixelWidth}{\pgfkeysvalueof{/tikz/optics/sensor line pixel width}}%
    \else%
      \pgfmathsetlengthmacro{\pixelWidth}{\pgfkeysvalueof{/tikz/optics/sensor line pixel width}*\objectWidth}%
    \fi%
  }

  \savedmacro\pixelHeight{%
    \pgfmathparse{(\objectHeight-2*\innerysep)/\pixelNumber}
    \edef\pixelHeight{\pgfmathresult pt}%
  }

  \savedanchor{\northeast}{
    \pgf@x=\objectWidth%
    \pgf@y=\objectHeight%
    \pgf@y=0.5\pgf@y%
    \pgf@x=0.5\pgf@x%
  }
  \anchor{north east}{\northeast}

  \savedanchor{\southwest}{
    \pgf@x=\objectWidth%
    \pgf@y=\objectHeight%
    \pgf@y=-0.5\pgf@y%
    \pgf@x=-0.5\pgf@x%
  }
  \anchor{south west}{\southwest}

  \anchor{north}{
    \center \pgf@xa=\pgf@x \pgf@ya=\pgf@y
    \northeast \pgf@xb=\pgf@x \pgf@yb=\pgf@y
    \pgf@x=\pgf@xa
    \pgf@y=\pgf@yb
  }

  \anchor{south}{
    \center \pgf@xa=\pgf@x \pgf@ya=\pgf@y
    \southwest \pgf@xb=\pgf@x \pgf@yb=\pgf@y
    \pgf@x=\pgf@xa
    \pgf@y=\pgf@yb
  }

  \anchor{east}{
    \center \pgf@xa=\pgf@x \pgf@ya=\pgf@y
    \northeast \pgf@xb=\pgf@x \pgf@yb=\pgf@y
    \pgf@x=\pgf@xb
    \pgf@y=\pgf@ya
  }

  \anchor{west}{
    \center \pgf@xa=\pgf@x \pgf@ya=\pgf@y
    \southwest \pgf@xb=\pgf@x \pgf@yb=\pgf@y
    \pgf@x=\pgf@xb
    \pgf@y=\pgf@ya
  }

  \anchor{north west}{
    \northeast \pgf@xa=\pgf@x \pgf@ya=\pgf@y
    \southwest \pgf@xb=\pgf@x \pgf@yb=\pgf@y
    \pgf@x=\pgf@xb
    \pgf@y=\pgf@ya
  }

  \anchor{south east}{
    \northeast \pgf@xa=\pgf@x \pgf@ya=\pgf@y
    \southwest \pgf@xb=\pgf@x \pgf@yb=\pgf@y
    \pgf@x=\pgf@xa
    \pgf@y=\pgf@yb
  }

  \expandafter\pgfutil@g@addto@macro\csname pgf@sh@s@sensor line\endcsname{%
    \c@pgf@counta\pixelNumber\relax%
    \pgfmathloop%
      \ifnum\c@pgf@counta>0\relax%
        \pgfutil@ifundefined{pgf@anchor@sensor line@pixel\space\the\c@pgf@counta\space north west}{%
        \expandafter\xdef\csname pgf@anchor@sensor line@pixel\space\the\c@pgf@counta\space north\endcsname{%
          \noexpand\northeast \noexpand\pgf@xa=\noexpand\pgf@x \noexpand\pgf@ya=\noexpand\pgf@y
          \noexpand\southwest \noexpand\pgf@xb=\noexpand\pgf@x \noexpand\pgf@yb=\noexpand\pgf@y
          \noexpand\pgf@x=\noexpand\pixelWidth
          \noexpand\pgf@x=.5\noexpand\pgf@x
          \noexpand\advance\noexpand\pgf@x by\noexpand\pgf@xb
          \noexpand\pgf@y=\noexpand\pgf@ya
          \noexpand\newdimen\noexpand\temp@y
          \noexpand\pgfmathsetlength\noexpand\temp@y{(\the\c@pgf@counta-\noexpand\pixelNumber)*\noexpand\pixelHeight-\noexpand\innerysep}
          \noexpand\advance\noexpand\pgf@y by\noexpand\temp@y
        }%
        \expandafter\xdef\csname pgf@anchor@sensor line@pixel\space\the\c@pgf@counta\space north east\endcsname{%
          \noexpand\northeast \noexpand\pgf@xa=\noexpand\pgf@x \noexpand\pgf@ya=\noexpand\pgf@y
          \noexpand\southwest \noexpand\pgf@xb=\noexpand\pgf@x \noexpand\pgf@yb=\noexpand\pgf@y
          \noexpand\pgf@x=\noexpand\pixelWidth
          \noexpand\advance\noexpand\pgf@x by\noexpand\pgf@xb
          \noexpand\pgf@y=\noexpand\pgf@ya
          \noexpand\newdimen\noexpand\temp@y
          \noexpand\pgfmathsetlength\noexpand\temp@y{(\the\c@pgf@counta-\noexpand\pixelNumber)*\noexpand\pixelHeight-\noexpand\innerysep}
          \noexpand\advance\noexpand\pgf@y by\noexpand\temp@y
        }%
        \expandafter\xdef\csname pgf@anchor@sensor line@pixel\space\the\c@pgf@counta\space north west\endcsname{%
          \noexpand\northeast \noexpand\pgf@xa=\noexpand\pgf@x \noexpand\pgf@ya=\noexpand\pgf@y
          \noexpand\southwest \noexpand\pgf@xb=\noexpand\pgf@x \noexpand\pgf@yb=\noexpand\pgf@y
          \noexpand\pgf@x=\noexpand\pgf@xb
          \noexpand\pgf@y=\noexpand\pgf@ya
          \noexpand\newdimen\noexpand\temp@y
          \noexpand\pgfmathsetlength\noexpand\temp@y{(\the\c@pgf@counta-\noexpand\pixelNumber)*\noexpand\pixelHeight-\noexpand\innerysep}
          \noexpand\advance\noexpand\pgf@y by\noexpand\temp@y
        }%
        \expandafter\xdef\csname pgf@anchor@sensor line@pixel\space\the\c@pgf@counta\space south\endcsname{%
          \noexpand\northeast \noexpand\pgf@xa=\noexpand\pgf@x \noexpand\pgf@ya=\noexpand\pgf@y
          \noexpand\southwest \noexpand\pgf@xb=\noexpand\pgf@x \noexpand\pgf@yb=\noexpand\pgf@y
          \noexpand\pgf@x=\noexpand\pixelWidth
          \noexpand\pgf@x=.5\noexpand\pgf@x
          \noexpand\advance\noexpand\pgf@x by\noexpand\pgf@xb
          \noexpand\pgf@y=\noexpand\pgf@ya
          \noexpand\newdimen\noexpand\temp@y
          \noexpand\pgfmathsetlength\noexpand\temp@y{(\the\c@pgf@counta-\noexpand\pixelNumber-1)*\noexpand\pixelHeight-\noexpand\innerysep}
          \noexpand\advance\noexpand\pgf@y by\noexpand\temp@y
        }%
        \expandafter\xdef\csname pgf@anchor@sensor line@pixel\space\the\c@pgf@counta\space south east\endcsname{%
          \noexpand\northeast \noexpand\pgf@xa=\noexpand\pgf@x \noexpand\pgf@ya=\noexpand\pgf@y
          \noexpand\southwest \noexpand\pgf@xb=\noexpand\pgf@x \noexpand\pgf@yb=\noexpand\pgf@y
          \noexpand\pgf@x=\noexpand\pixelWidth
          \noexpand\advance\noexpand\pgf@x by\noexpand\pgf@xb
          \noexpand\pgf@y=\noexpand\pgf@ya
          \noexpand\newdimen\noexpand\temp@y
          \noexpand\pgfmathsetlength\noexpand\temp@y{(\the\c@pgf@counta-\noexpand\pixelNumber-1)*\noexpand\pixelHeight-\noexpand\innerysep}
          \noexpand\advance\noexpand\pgf@y by\noexpand\temp@y
        }%
        \expandafter\xdef\csname pgf@anchor@sensor line@pixel\space\the\c@pgf@counta\space south west\endcsname{%
          \noexpand\northeast \noexpand\pgf@xa=\noexpand\pgf@x \noexpand\pgf@ya=\noexpand\pgf@y
          \noexpand\southwest \noexpand\pgf@xb=\noexpand\pgf@x \noexpand\pgf@yb=\noexpand\pgf@y
          \noexpand\pgf@x=\noexpand\pgf@xb
          \noexpand\pgf@y=\noexpand\pgf@ya
          \noexpand\newdimen\noexpand\temp@y
          \noexpand\pgfmathsetlength\noexpand\temp@y{(\the\c@pgf@counta-\noexpand\pixelNumber-1)*\noexpand\pixelHeight-\noexpand\innerysep}
          \noexpand\advance\noexpand\pgf@y by\noexpand\temp@y
        }%
        \expandafter\xdef\csname pgf@anchor@sensor line@pixel\space\the\c@pgf@counta\space east\endcsname{%
          \noexpand\northeast \noexpand\pgf@xa=\noexpand\pgf@x \noexpand\pgf@ya=\noexpand\pgf@y
          \noexpand\southwest \noexpand\pgf@xb=\noexpand\pgf@x \noexpand\pgf@yb=\noexpand\pgf@y
          \noexpand\pgf@x=\noexpand\pixelWidth
          \noexpand\advance\noexpand\pgf@x by\noexpand\pgf@xb
          \noexpand\pgf@y=\noexpand\pgf@ya
          \noexpand\newdimen\noexpand\temp@y
          \noexpand\pgfmathsetlength\noexpand\temp@y{(\the\c@pgf@counta-\noexpand\pixelNumber-0.5)*\noexpand\pixelHeight-\noexpand\innerysep}
          \noexpand\advance\noexpand\pgf@y by\noexpand\temp@y
        }%
        \expandafter\xdef\csname pgf@anchor@sensor line@pixel\space\the\c@pgf@counta\space center\endcsname{%
          \noexpand\northeast \noexpand\pgf@xa=\noexpand\pgf@x \noexpand\pgf@ya=\noexpand\pgf@y
          \noexpand\southwest \noexpand\pgf@xb=\noexpand\pgf@x \noexpand\pgf@yb=\noexpand\pgf@y
          \noexpand\pgf@x=\noexpand\pixelWidth
          \noexpand\pgf@x=.5\noexpand\pgf@x
          \noexpand\advance\noexpand\pgf@x by\noexpand\pgf@xb
          \noexpand\pgf@y=\noexpand\pgf@ya
          \noexpand\newdimen\noexpand\temp@y
          \noexpand\pgfmathsetlength\noexpand\temp@y{(\the\c@pgf@counta-\noexpand\pixelNumber-0.5)*\noexpand\pixelHeight-\noexpand\innerysep}
          \noexpand\advance\noexpand\pgf@y by\noexpand\temp@y
        }%
        \expandafter\xdef\csname pgf@anchor@sensor line@pixel\space\the\c@pgf@counta\space west\endcsname{%
          \noexpand\northeast \noexpand\pgf@xa=\noexpand\pgf@x \noexpand\pgf@ya=\noexpand\pgf@y
          \noexpand\southwest \noexpand\pgf@xb=\noexpand\pgf@x \noexpand\pgf@yb=\noexpand\pgf@y
          \noexpand\pgf@x=\noexpand\pgf@xb
          \noexpand\pgf@y=\noexpand\pgf@ya
          \noexpand\newdimen\noexpand\temp@y
          \noexpand\pgfmathsetlength\noexpand\temp@y{(\the\c@pgf@counta-\noexpand\pixelNumber-0.5)*\noexpand\pixelHeight-\noexpand\innerysep}
          \noexpand\advance\noexpand\pgf@y by\noexpand\temp@y
        }%
      }{\c@pgf@counta0\relax}%
      \advance\c@pgf@counta-1\relax%
    \repeatpgfmathloop%
  }%

  \inheritanchorborder[from=rectangle]

  \backgroundpath
  {
    \pgfpathrectanglecorners{\northeast}{\southwest}

    \c@pgf@counta\pixelNumber\relax%
    \pgfmathloop%
      \ifnum\c@pgf@counta>0\relax%
        \newdimen\pixel@northeast@x
        \newdimen\pixel@northeast@y
        \newdimen\pixel@southwest@x
        \newdimen\pixel@southwest@y
        \northeast \pgf@xa=\pgf@x \pgf@ya=\pgf@y
        \southwest \pgf@xb=\pgf@x \pgf@yb=\pgf@y
        \pixel@northeast@x=\pgf@xb
        \advance\pixel@northeast@x by\pixelWidth
        \pixel@northeast@y=\pgf@ya
        \newdimen\temp@y
        \pgfmathsetlength\temp@y{(\the\c@pgf@counta-\pixelNumber)*\pixelHeight-\innerysep}
        \advance\pixel@northeast@y by\noexpand\temp@y
        \pixel@southwest@x=\pgf@xb
        \pixel@southwest@y=\pgf@ya
        \newdimen\temp@y
        \pgfmathsetlength\temp@y{(\the\c@pgf@counta-\pixelNumber-1)*\pixelHeight-\innerysep}
        \advance\pixel@southwest@y by\temp@y
        \pgfpathrectanglecorners{\pgfpoint{\pixel@northeast@x}{\pixel@northeast@y}}{\pgfpoint{\pixel@southwest@x}{\pixel@southwest@y}}
      \advance\c@pgf@counta-1\relax%
    \repeatpgfmathloop%
  }
}

\pgfkeys{/tikz/optics/screen/.style={thin optics element, very thick}}
\pgfkeys{/tikz/optics/diffraction grating/.style={thin optics element, /tikz/optics/cheating dash={on 4pt off 2pt}}}
\pgfkeys{/tikz/optics/grid/.style={thin optics element, /tikz/optics/cheating dash={on 4pt off 2pt}, ultra thick}}
\pgfkeys{/tikz/optics/semi-transparent mirror/.style={thin optics element, densely dotted, thick}}
\pgfkeys{/tikz/optics/diaphragm/.style={slit,/tikz/optics/slit height=0.4}}

\pgfkeys{/tikz/optics/generic lamp/.style={shape=generic optics io,optics,draw}}
\pgfkeys{/tikz/optics/generic sensor/.style={shape=generic optics io,optics,io orientation=rtl,draw, io body height=1cm, io body aspect ratio=0.5, io aperture width=0.15}}
\pgfkeys{/tikz/optics/halogen lamp/.style={/tikz/optics/generic lamp, io body height=0.75cm, io body aspect ratio=2,io aperture width=0.33,io aperture height=0.66,io aperture shift=0}}
\pgfkeys{/tikz/optics/spectral lamp/.style={/tikz/optics/generic lamp, io body height=2.25cm, io body aspect ratio=2/3,io aperture width=0.11,io aperture height=0.22,io aperture shift=0.25,/tikz/optics/io multiline}}
\pgfkeys{/tikz/optics/laser/.style={/tikz/optics/generic lamp, io body height=0.5cm, io body aspect ratio=3,io aperture width=0.22,io aperture height=0.5,io aperture shift=0}}
\pgfkeys{/tikz/optics/laser'/.style={/tikz/optics/generic lamp, io body height=0.5cm, io body aspect ratio=3,io aperture width=0,io aperture height=0.5,io aperture shift=0}}
\pgfkeys{/tikz/optics/beam splitter/.style={shape=polarizer, optics, draw, object height=1cm, object aspect ratio=1}}

\tikzset{
    /tikz/optics/cheating dash/.code args={on #1 off #2}{
        \csname tikz@addoption\endcsname{%
            \pgfgetpath\currentpath%
            \pgfprocessround{\currentpath}{\currentpath}%
            \csname pgf@decorate@parsesoftpath\endcsname{\currentpath}{\currentpath}%
            \pgfmathparse{\csname pgf@decorate@totalpathlength\endcsname-#1}\let\rest=\pgfmathresult%
            \pgfmathparse{#1+#2}\let\onoff=\pgfmathresult%
            \pgfmathparse{max(floor(\rest/\onoff), 1)}\let\nfullonoff=\pgfmathresult%
            \pgfmathparse{max((\rest-\onoff*\nfullonoff)/\nfullonoff+#2, #2)}\let\offexpand=\pgfmathresult%
            \pgfsetdash{{#1}{\offexpand}}{0pt}}%
    }
}

\pgfkeys{/tikz/optics/io multiline/.code={\newdimen\tmplen\pgfmathsetlength{\tmplen}{\pgfkeysvalueof{/tikz/optics/io body aspect ratio}*\pgfkeysvalueof{/tikz/optics/io body height}}\tikzset{text width=\the\tmplen,align=center}}}

\pgfkeys{/tikz/optics/mark point/.style={optics/mark a cross}}

\pgfkeys{/tikz/optics/draw focal points/.style={append after command={
    \pgfextra{
      \begin{pgfinterruptpath}
          \node[/tikz/optics/mark point,#1] at (\tikzlastnode.west focal point) {};
          \node[/tikz/optics/mark point,#1] at (\tikzlastnode.east focal point) {};
      \end{pgfinterruptpath}
    }
  }
}}

\pgfkeys{/tikz/optics/draw mirror focus/.style={append after command={
    \pgfextra{
      \begin{pgfinterruptpath}
          \node[/tikz/optics/mark point,#1] at (\tikzlastnode.focus) {};
      \end{pgfinterruptpath}
    }
  }
}}

\pgfkeys{/tikz/optics/draw mirror center/.style={append after command={
    \pgfextra{
      \begin{pgfinterruptpath}
          \node[/tikz/optics/mark point,#1] at (\tikzlastnode.mirror center) {};
      \end{pgfinterruptpath}
    }
  }
}}

\pgfkeys{/tikz/optics/mark a cross/.style={cross out,draw,inner sep=0pt,minimum width=2pt,minimum height=2pt}}

\tikzset{
  put coordinate/.style args={#1 at #2}{decoration={markings, mark=at position #2 with {\node[coordinate] (#1) {};}},postaction={decorate}}
}

\tikzset{/tikz/put arrow/.cd,
  pos/.initial=0.5,
  at/.style={pos=#1},
  pos var/.initial={},
  style/.initial={},
  style var/.initial={},
  postaction style/.initial={}, 
  arrow macro/.initial={arrow},
  arrow macro var/.initial={},
  arrow type/.initial={>},
  arrow type var/.initial={},
  reversed/.style={arrow macro=arrowreversed},
  arrow/.style={arrow type=#1},
  arrow'/.style={arrow type=#1, reversed},
  every arrow/.style={},
}

\tikzset{put arrow/.code={
  \tikzset{/tikz/put arrow/.cd, #1}
  \tikzset{/tikz/put arrow/pos var/.expand once={\pgfkeysvalueof{/tikz/put arrow/pos}}}
  \tikzset{/tikz/put arrow/style var/.expanded={\pgfkeysvalueof{/tikz/put arrow/style}}}
  \tikzset{/tikz/put arrow/arrow macro var/.expand once={\pgfkeysvalueof{/tikz/put arrow/arrow macro}}}
  \tikzset{/tikz/put arrow/arrow type var/.expand once={\pgfkeysvalueof{/tikz/put arrow/arrow type}}}
  \tikzset{/tikz/put arrow/ordered draw key/.expanded=%
      {\pgfkeysvalueof{/tikz/put arrow/pos var}}%
      {\pgfkeysvalueof{/tikz/put arrow/style var}}%
      {\pgfkeysvalueof{/tikz/put arrow/arrow macro var}}%
      {\pgfkeysvalueof{/tikz/put arrow/arrow type var}}%
      {\pgfkeysvalueof{/tikz/put arrow/postaction style}}%
  }
  \tikzset{/tikz/put arrow/pos=0.5} 
  \tikzset{/tikz/put arrow/style={}}
  \tikzset{/tikz/put arrow/arrow macro={arrow}}
  \tikzset{/tikz/put arrow/arrow type={>}}
}}

\tikzset{/tikz/put arrow/ordered draw key/.code n args={5}{
  \tikzset{postaction={#5,decorate, decoration={markings, mark=at position #1 with {\csname #3\endcsname[/tikz/put arrow/every arrow,#2]{#4}};}}}
}}

\tikzset{/tikz/optics/multiple ray arrow/.cd,
  n/.initial=1,
  n var/.initial=1,
  set n/.code={\pgfsetarrowoptions{multiple ray arrow}{#1}},
}

\pgfkeys{
  /tikz/optics/.cd, 
  use ray arrow >/.code 2 args={
    \pgfsetarrowoptions{ray arrow@length}{4pt}
    \pgfsetarrowoptions{ray arrow@angle}{50}
    \tikzset{/tikz/put arrow/postaction style/.expanded={/tikz/optics/multiple ray arrow/set n=#1}}
    \tikzset{/tikz/put arrow/.expanded={arrow={multiple ray arrow}, #2}}
  },
  use ray arrow </.code 2 args={
    \pgfsetarrowoptions{ray arrow@length}{4pt}
    \pgfsetarrowoptions{ray arrow@angle}{50}
    \tikzset{/tikz/put arrow/postaction style/.expanded={/tikz/optics/multiple ray arrow/set n=#1}}
    \tikzset{/tikz/put arrow/.expanded={arrow'={multiple ray arrow}, #2}}
  },
  ->n-/.code={
    \tikzset{/tikz/optics/multiple ray arrow/.cd, .collect unknowns,%
      #1,
      unknown options/.get = \arrowkeys}
    \tikzset{/tikz/optics/multiple ray arrow/n var/.expand once={\pgfkeysvalueof{/tikz/optics/multiple ray arrow/n}}}
    \tikzset{/tikz/optics/use ray arrow >={\pgfkeysvalueof{/tikz/optics/multiple ray arrow/n var}}{\arrowkeys}}
  },
  -<n-/.code={
    \tikzset{/tikz/optics/multiple ray arrow/.cd, .collect unknowns,%
      #1,
      unknown options/.get = \arrowkeys}
    \tikzset{/tikz/optics/multiple ray arrow/n var/.expand once={\pgfkeysvalueof{/tikz/optics/multiple ray arrow/n}}}
    \tikzset{/tikz/optics/use ray arrow <={\pgfkeysvalueof{/tikz/optics/multiple ray arrow/n var}}{\arrowkeys}}
  },
  ->-/.style={/tikz/optics/->n-={n=1, #1}},
  -<-/.style={/tikz/optics/-<n-={n=1, #1}},
  ->>-/.style={/tikz/optics/->n-={n=2, #1}},
  -<<-/.style={/tikz/optics/-<n-={n=2, #1}},
  ->>>-/.style={/tikz/optics/->n-={n=3, #1}},
  -<<<-/.style={/tikz/optics/-<n-={n=3, #1}},
  ->>>>-/.style={/tikz/optics/->n-={n=4, #1}},
  -<<<<-/.style={/tikz/optics/-<n-={n=4, #1}},
}

\def\dimarrow@short@position{0}
\newif\ifdimarrow@nearstart
\dimarrow@nearstarttrue
\tikzset{%
  /tikz/dim arrow/.code={\tikzset{draw,/tikz/dim arrow/draw dim arrow}\pgfkeys{/tikz/dim arrow/.cd,#1}},
  /tikz/dim arrow'/.code={\pgfkeysgetvalue{/tikz/dim arrow/raise}{\tmp@tdar}\tikzset{draw,/tikz/dim arrow/draw dim arrow,/tikz/dim arrow/raise=-\tmp@tdar}\pgfkeys{/tikz/dim arrow/.cd,#1}},
  /tikz/short dim arrow/.code={\tikzset{draw,/tikz/dim arrow/draw short dim arrow}\pgfkeys{/tikz/dim arrow/.cd,#1}},
  /tikz/short dim arrow'/.code={\pgfkeysgetvalue{/tikz/dim arrow/raise}{\tmp@tdar}\tikzset{draw,/tikz/dim arrow/draw short dim arrow,/tikz/dim arrow/raise=-\tmp@tdar}\pgfkeys{/tikz/dim arrow/.cd,#1}},
  /tikz/dim arrow/.cd,
  raise/.initial={0.5cm},
  no raise/.style={raise=0},
  label/.code={\pgfkeys{/tikz/dim arrow/label text=#1}},
  label'/.code={\pgfkeys{/tikz/dim arrow/label text=#1,/tikz/dim arrow/label style/.append style={swap},}},
  label text/.initial={},
  label style/.style={},
  label near start/.code={\def\dimarrow@short@position{0}}, 
  label near middle/.code={\def\dimarrow@short@position{2}}, 
  label near end/.code={\def\dimarrow@short@position{1}}, 
  arrow length/.initial={5mm}, 
  draw short dim arrow/.style={to path={\pgfextra{%
    \let\tikz@mode@save=\tikz@mode%
        \let\tikz@options@save=\tikz@options%
    \newdimen\labelTotalRaise
    \pgfmathsetlength\labelTotalRaise{\pgfkeysvalueof{/tikz/dim arrow/raise}}
        \pgfinterruptpath
        \draw[>=technical,->|] \pgfextra{\let\tikz@mode=\tikz@mode@save\let\tikz@options=\tikz@options@save}
    let
        \p1=($(\tikztostart)!\pgfkeysvalueof{/tikz/dim arrow/raise}!90:(\tikztotarget)$),
        \p2=($(\tikztotarget)!\pgfkeysvalueof{/tikz/dim arrow/raise}!-90:(\tikztostart)$)
        in ($(\p1)!-\pgfkeysvalueof{/tikz/dim arrow/arrow length}!(\p2)$) -- ($(\p1)!0!(\p2)$);
    \draw[>=technical,->|] \pgfextra{\let\tikz@mode=\tikz@mode@save\let\tikz@options=\tikz@options@save}
    let
        \p1=($(\tikztostart)!\pgfkeysvalueof{/tikz/dim arrow/raise}!90:(\tikztotarget)$),
        \p2=($(\tikztotarget)!\pgfkeysvalueof{/tikz/dim arrow/raise}!-90:(\tikztostart)$)
        in ($(\p2)!-\pgfkeysvalueof{/tikz/dim arrow/arrow length}!(\p1)$) -- ($(\p2)!0!(\p1)$);
    \ifnum\dimarrow@short@position=0
    \path let 
        \p1=($(\tikztostart)!\labelTotalRaise!90:(\tikztotarget)$),
        \p2=($(\tikztotarget)!\labelTotalRaise!-90:(\tikztostart)$)
        in let
    \p3=($(\p1)!-1*\pgfkeysvalueof{/tikz/dim arrow/arrow length}!(\p2)$),
    \p4=($(\p1)!-0*\pgfkeysvalueof{/tikz/dim arrow/arrow length}!(\p2)$)
    in
    (\p3) -- (\p4) node[pos=0.5,auto=left,/tikz/dim arrow/label style] {\pgfkeysvalueof{/tikz/dim arrow/label text}};
  \fi
    \ifnum\dimarrow@short@position=1
      \path let 
        \p1=($(\tikztostart)!\labelTotalRaise!90:(\tikztotarget)$),
        \p2=($(\tikztotarget)!\labelTotalRaise!-90:(\tikztostart)$)
        in let
    \p3=($(\p2)!-1*\pgfkeysvalueof{/tikz/dim arrow/arrow length}!(\p1)$),
    \p4=($(\p2)!-0*\pgfkeysvalueof{/tikz/dim arrow/arrow length}!(\p1)$)
    in
    (\p4) -- (\p3) node[pos=0.5,auto=left,/tikz/dim arrow/label style] {\pgfkeysvalueof{/tikz/dim arrow/label text}};
    \fi
  \ifnum\dimarrow@short@position=2
    \path let 
        \p1=($(\tikztostart)!\labelTotalRaise!90:(\tikztotarget)$),
        \p2=($(\tikztotarget)!\labelTotalRaise!-90:(\tikztostart)$)
    in
    (\p1) -- (\p2) node[pos=0.5,/tikz/dim arrow/label style] {\pgfkeysvalueof{/tikz/dim arrow/label text}};
  \fi
        \endpgfinterruptpath
      }(\tikztostart) (\tikztotarget) \tikztonodes
    }
  },
  draw dim arrow/.style={to path={\pgfextra{%
    \let\tikz@mode@save=\tikz@mode%
        \let\tikz@options@save=\tikz@options%
    \newdimen\labelTotalRaise
    \pgfmathsetlength\labelTotalRaise{\pgfkeysvalueof{/tikz/dim arrow/raise}}
        \pgfinterruptpath
        \draw[>=technical,|<->|] \pgfextra{\let\tikz@mode=\tikz@mode@save\let\tikz@options=\tikz@options@save}
    let 
        \p1=($(\tikztostart)!\pgfkeysvalueof{/tikz/dim arrow/raise}!90:(\tikztotarget)$),
        \p2=($(\tikztotarget)!\pgfkeysvalueof{/tikz/dim arrow/raise}!-90:(\tikztostart)$)
        in (\p1) -- (\p2);
    \path let 
        \p1=($(\tikztostart)!\labelTotalRaise!90:(\tikztotarget)$),
        \p2=($(\tikztotarget)!\labelTotalRaise!-90:(\tikztostart)$)
        in (\p1) -- (\p2) node[pos=0.5,auto=left,/tikz/dim arrow/label style] {\pgfkeysvalueof{/tikz/dim arrow/label text}};
        \endpgfinterruptpath
      }(\tikztostart) (\tikztotarget) \tikztonodes
    }
  },
}

\pgfsetarrowoptions{lens arrow@length}{6pt}
\pgfsetarrowoptions{lens arrow@angle}{50}
\pgfarrowsdeclare{lens arrow}{lens arrow}
{
   \pgfarrowsleftextend{0pt}
   \pgfarrowsrightextend{0pt}
}
{
  \pgfsetroundcap
  \pgfsetmiterjoin
  \pgfmathsetlength{\pgfutil@tempdimb}{\pgfgetarrowoptions{lens arrow@length}*sin(\pgfgetarrowoptions{lens arrow@angle}/2)}    
  \def\arrow@origin{\pgfpoint{0pt}{0pt}}
  \pgfutil@tempdima=\pgfgetarrowoptions{lens arrow@length}%
  \pgfmathsetmacro{\tmp@lens@angle}{90+\pgfgetarrowoptions{lens arrow@angle}}
  \pgfmathsetmacro{\tmp@lens@anglediv}{\pgfgetarrowoptions{lens arrow@angle}/2}
  \advance\pgfutil@tempdima by -1.5\pgflinewidth%
  \pgfmathsetlength{\pgfutil@tempdima}{\pgfutil@tempdima/cos(\pgfgetarrowoptions{lens arrow@angle}/2)}    
  \pgfpathmoveto{\pgfpointadd{\arrow@origin}{\pgfqpointpolar{\tmp@lens@angle}{\pgfutil@tempdima}}}
  \pgfpathlineto{\arrow@origin}
  \pgfpathlineto{\pgfpointadd{\arrow@origin}{\pgfqpointpolar{-\tmp@lens@angle}{\pgfutil@tempdima}}}
  \pgfusepathqstroke
}
\pgfarrowsdeclarereversed{lens arrow reversed}{lens arrow reversed}{lens arrow}{lens arrow}

\pgfsetarrowoptions{ray arrow@length}{4pt}
\pgfsetarrowoptions{ray arrow@angle}{45}

\makeatletter

\pgfsetarrowoptions{multiple ray arrow}{0}
\pgfarrowsdeclare{multiple ray arrow}{multiple ray arrow}
{
    \pgfarrowsleftextend{0pt}
    \pgfarrowsrightextend{0pt}
}
{
  \pgfsetroundcap
  \pgfsetmiterjoin
  \pgfutil@tempdima=\pgfgetarrowoptions{ray arrow@length}%
  \pgfmathsetmacro{\tmp@ray@angle}{90+\pgfgetarrowoptions{ray arrow@angle}}
  \pgfmathsetmacro{\tmp@ray@anglediv}{\pgfgetarrowoptions{ray arrow@angle}/2}
  \advance\pgfutil@tempdima by -1.5\pgflinewidth%
  \pgfmathsetlength{\pgfutil@tempdima}{\pgfutil@tempdima/cos(\pgfgetarrowoptions{ray arrow@angle}/2)}
  \foreach \i in {1,...,\pgfgetarrowoptions{multiple ray arrow}}
  {
    \pgfmathsetlength{\pgfutil@tempdimb}{(2*\i-\pgfgetarrowoptions{multiple ray arrow})*\pgfgetarrowoptions{ray arrow@length}*sin(\pgfgetarrowoptions{ray arrow@angle}/2)}
    \def\arrow@origin{\pgfpoint{\pgfutil@tempdimb}{0pt}}
    \pgfpathmoveto{\pgfpointadd{\arrow@origin}{\pgfqpointpolar{\tmp@ray@angle}{\pgfutil@tempdima}}}
    \pgfpathlineto{\arrow@origin}
    \pgfpathlineto{\pgfpointadd{\arrow@origin}{\pgfqpointpolar{-\tmp@ray@angle}{\pgfutil@tempdima}}}
    \pgfusepathqstroke
  }
}

\makeatletter
\pgfarrowsdeclare{technical}{technical}
{%
  \pgfutil@tempdima=0.48pt%
  \pgfutil@tempdimb=\pgflinewidth%
  \ifdim\pgfinnerlinewidth>0pt%
    \pgfmathsetlength\pgfutil@tempdimb{.6\pgflinewidth-.4*\pgfinnerlinewidth}%
  \fi%
  \advance\pgfutil@tempdima by.3\pgfutil@tempdimb%
  \pgfarrowsleftextend{+-3\pgfutil@tempdima}%
  \pgfarrowsrightextend{+8\pgfutil@tempdima}%
}
{%
  \pgfutil@tempdima=0.48pt%
  \pgfutil@tempdimb=\pgflinewidth%
  \ifdim\pgfinnerlinewidth>0pt%
    \pgfmathsetlength\pgfutil@tempdimb{.6\pgflinewidth-.4*\pgfinnerlinewidth}%
  \fi%
  \advance\pgfutil@tempdima by.3\pgfutil@tempdimb%
  \pgfpathmoveto{\pgfqpoint{8\pgfutil@tempdima}{0pt}}%
  \pgfpathlineto{\pgfqpoint{-3\pgfutil@tempdima}{3\pgfutil@tempdima}}%
  \pgfpathlineto{\pgfpointorigin}%
  \pgfpathlineto{\pgfqpoint{-3\pgfutil@tempdima}{-3\pgfutil@tempdima}}%
  \pgfusepathqfill%
}

\pgfarrowsdeclare{technical reversed}{technical reversed}
{%
  \pgfutil@tempdima=0.48pt%
  \pgfutil@tempdimb=\pgflinewidth%
  \ifdim\pgfinnerlinewidth>0pt%
    \pgfmathsetlength\pgfutil@tempdimb{.6\pgflinewidth-.4*\pgfinnerlinewidth}%
  \fi%
  \advance\pgfutil@tempdima by.3\pgfutil@tempdimb%
  \pgfarrowsleftextend{-8\pgfutil@tempdima}
  \pgfarrowsrightextend{-8\pgfutil@tempdima}
}
{%
  \pgfutil@tempdima=0.48pt%
  \pgfutil@tempdimb=\pgflinewidth%
  \ifdim\pgfinnerlinewidth>0pt%
    \pgfmathsetlength\pgfutil@tempdimb{.6\pgflinewidth-.4*\pgfinnerlinewidth}%
  \fi%
  \advance\pgfutil@tempdima by.3\pgfutil@tempdimb%
  \pgfpathmoveto{\pgfqpoint{-8\pgfutil@tempdima}{0pt}}%
  \pgfpathlineto{\pgfqpoint{3\pgfutil@tempdima}{3\pgfutil@tempdima}}%
  \pgfpathlineto{\pgfpointorigin}%
  \pgfpathlineto{\pgfqpoint{3\pgfutil@tempdima}{-3\pgfutil@tempdima}}%
  \pgfusepathqfill%
}

\makeatother

\allowdisplaybreaks 

\usepackage{mathtools}
\usepackage{xfrac}
\usepackage{bm}

\usepackage{makecell}
\usepackage{multirow}
\usepackage{scrextend} 

\usepackage[colorlinks]{hyperref}

\usepackage{newunicodechar}

\newunicodechar{α}{\alpha}
\newunicodechar{β}{\beta}
\newunicodechar{κ}{\kappa}  
\newunicodechar{λ}{\lambda} 
\newunicodechar{Λ}{\Lambda} 
\newunicodechar{υ}{\upsilon} 
\newunicodechar{ϒ}{\Upsilon}
\newunicodechar{π}{\pi}
\newunicodechar{Π}{\Pi}
\newunicodechar{Φ}{\Phi}
\newunicodechar{Ψ}{\Psi}
\newunicodechar{ψ}{\psi}

\newunicodechar{±}{\pm}
\newunicodechar{⋅}{\cdot}
\newunicodechar{×}{\times}
\newunicodechar{⊗}{\otimes}
\newunicodechar{≠}{\neq}
\newunicodechar{≤}{\leq}
\newunicodechar{≥}{\geq}
\newunicodechar{∈}{\in}

\newunicodechar{∀}{\forall}
\newunicodechar{∃}{\exists}
\newunicodechar{√}{\sqrt}
\newunicodechar{∑}{\sum}
\newunicodechar{⋯}{\cdots}
\newunicodechar{⌈}{\lceil}
\newunicodechar{⌉}{\rceil}

\newunicodechar{⁺}{^+}
\newunicodechar{⁻}{^-}
\newunicodechar{ⁿ}{^n}
\newunicodechar{¹}{^1}
\newunicodechar{²}{^2}
\newunicodechar{ₑ}{_e}
\newunicodechar{₁}{_1}
\newunicodechar{₂}{_2}
\newunicodechar{₃}{_3}
\newunicodechar{₄}{_4}

\newunicodechar{←}{\leftarrow}
\newunicodechar{↑}{\uparrow}
\newunicodechar{↓}{\downarrow}
\newunicodechar{→}{\rightarrow}

\newunicodechar{½}{\tfrac12}

\usepackage{physics}

\newcommand{\eg}{\emph{e.g.\@}}
\newcommand{\etal}{\emph{et al.\@}}
\newcommand{\ie}{\emph{i.e.\@}}
\newcommand{\twlog}{\emph{w.l.o.g.\@}} 

\DeclarePairedDelimiter{\ceil}{⌈}{⌉}
\DeclarePairedDelimiter{\floor}{\lfloor}{\rfloor}
\DeclarePairedDelimiter{\pars}{(}{)}

\newcommand{\e}{\mathrm e} 
\newcommand{\Prb}[1][]{\mathcal P_{#1}}
\newcommand{\PrbSu}{\Prb[\text{succ}]}
\DeclareMathOperator{\diag}{diag}

\begin{document}

\title{Investigating the optimality of ancilla-assisted linear optical Bell measurements}

\author{Andrea Olivo}
\email{andrea.olivo@u-psud.fr}
\affiliation{Inria, Paris, France}
\affiliation{Laboratoire Aimé Cotton, CNRS, Université Paris-Sud, ENS Cachan,
	Université Paris-Saclay, 91405 Orsay Cedex, France}

\author{Frédéric Grosshans}
\email{frederic.grosshans@u-psud.fr}
\affiliation{Laboratoire Aimé Cotton, CNRS, Université Paris-Sud, ENS Cachan,
	Université Paris-Saclay, 91405 Orsay Cedex, France}

\date{\today}

\begin{abstract}
In the last decade Grice~\cite{Grice2011ArbitrarilyElements} and Ewert and van Loock~\cite{Ewert20143Ancillae} 
found linear optical networks achieving near-unit efficiency unambiguous Bell state discrimination, 
when fed with increasingly complex ancillary states.
However, except for the vacuum ancilla case~\cite{Calsamiglia2001MaximumAnalyzer}, 
the optimality of these schemes is unknown.
Here, the optimality of these networks is investigated through analytical and numerical means.
We show an analytical upper bound to the success probability for interferometers that preserve the polarization of the input photons, saturated by both Grice's and Evert-van Loock's strategies.
Furthermore, such an upper bound links the complexity of their ancilla states with the scaling of their performance.
We also show a computer-aided approach to the optimization of such measurement schemes for generic interferometers, by simulating an optical network supplied with various kinds of ancillary input states.
We numerically confirms the optimality of known small schemes.
We use both methods to investigate other ancilla states, some of them never studied before.
\end{abstract}

\maketitle

\section{Introduction}\label{sec:level1}

Due to its experimental and theoretical simplicity, 
linear quantum optics has proved to be a promising route for the early implementation of 
important quantum communication protocols \cite{Braunstein1995MeasurementTeleportation}---including 
quantum teleportation~\cite{Bennett1993TeleportingChannels,Lee2012Bell-stateEntanglement,Bouwmeester1997ExperimentalTeleportation}, dense coding~\cite{Bennett1992CommunicationStates,Mattle1996DenseCommunication} and entanglement swapping~\cite{Zukowski1993Event-ready-detectorsSwapping,Pan1998ExperimentalInteracted}.
An essential step in these protocols is the \textit{Bell measurement} (BM), 
a projective measurement onto a basis of two-qubit maximally entangled states, the Bell states. 
In order to enable the common scenario where losses can be tolerated but not errors, in the following we will be
concerned with \textit{unambiguous} BM, \ie{} a measurement whose outcome is never wrong, but is sometimes inconclusive.

Lütkenhaus \etal{} \cite{Lutkenhaus1999BellTeleportation} showed long ago the 
impossibility of a linear optical perfect Bell measurement for dual-rail photonic qubits.
In a following result, Calsamglia and Lütkenhaus bound the success probability $\PrbSu$ of the no-ancilla case to
50\%~\cite{Calsamiglia2001MaximumAnalyzer}, thus proving the optimality of the already known 
Braunstein--Mann scheme~\cite{Braunstein1995MeasurementTeleportation}.
Recently, Grice~\cite{Grice2011ArbitrarilyElements}, followed by Ewert and van Loock~\cite{Ewert20143Ancillae}
showed how to overcome this bound by supplying the network with ancillary states.
They showed how to attain a success probability of $\PrbSu=3/4$ with reasonable ancillary states, and 
how to increase this probability to values arbitrarily close to 1, by using increasingly complex
and entangled ancillæ.
Other ways around the 50\% barrier include feed-forward techniques from linear optical quantum computation~\cite{Knill2001AOptics,Kok2007LinearQubits}, 
squeezing operations~\cite{Zaidi2013BeatingMeasurements}, Kerr nonlinearities~\cite{Vitali2000CompleteNonlinearity}, 
entangled coherent states, hybrid entanglement~\cite{Lee2012Bell-stateEntanglement,Lee2013Near-deterministicQubits} or encoded qubits \cite{Lee2015NearlyProcessing}.
While some of the above techniques may in principle
be used to realize a perfect BM, each one has its own disadvantages and present different experimental challenges in their implementation.
For linear optics the difficulties are concentrated in the preparation of complex ancillary states;
this is partly compensated by the simplicity of interferometers.

Our main motivation in this work is to investigate the optimality of ancilla-assisted linear optical schemes:
what is the highest possible $\PrbSu$ for a given ancilla? Which are the “simplest” states to achieve a given
value for $\PrbSu$? Very recently, and independently from us, Smith and Kaplan \cite{Smith2018ApproachingPhotons} tackled a 
similar problem, optimizing the mutual information of a Bell measurement using single photon ancillæ. 

In  Section~\ref{sec:UppBPolInd}, we first present an analytical upper bound to the success probability of 
ancilla-assisted BM for polarization-preserving interferometers. Specifically, this bound is saturated
by the Grice scheme \cite{Grice2011ArbitrarilyElements}.
In Section~\ref{sec:prog}, we expose a linear optical network optimizer based on symbolic and numerical computations,
built in order to maximize the success probability of unambiguous BM for a given ancillary state and to argue about 
its optimality.
This program is available as supplementary material to this article~\cite{Olivo2018BellOptimizer}.
In Section~\ref{sec:results}, we discuss the results obtained with the two approaches above with different
kinds of ancillæ and compare them to previously known results 
\cite{Braunstein1995MeasurementTeleportation, Calsamiglia2001MaximumAnalyzer, Grice2011ArbitrarilyElements,
  Ewert20143Ancillae,Smith2018ApproachingPhotons}. 
To our knowledge, some ancillæ studied here were never employed for this task.
We also discuss some of the hidden symmetries of the problem at hand, 
some of which we exploit in order to lessen the amount of computation needed.
Finally, we conclude this work in Section~\ref{sec:Conclusion}.

\section{Analytical upper bound for polarization-preserving interferometers}\label{sec:UppBPolInd}
A desirable goal is to find an unambiguous Bell measurement using linear optics with the best possible 
success probability $\PrbSu$ and the simplest possible ancillary state.
In this section, we present an analytical upper bound for the success probability $\PrbSu$ of such unambiguous
discrimination schemes using ancillary states and polarization preserving linear optics.
This restriction to polarization preserving interferometers---\ie\ 
networks described by a block-diagonal unitary
$U=\diag(U_h,U_v)$, with $U_h$ (resp.\@ $U_v$) acting on the horizontally (resp.\@ vertically) polarized modes---%
is not motivated by experimental realities but is an artificial consequence of the proof technique.
Specifically, this restriction will not be enforced in the computer aided optimization of Section~\ref{sec:prog}.
However, the previously known schemes
\cite{Braunstein1995MeasurementTeleportation,Grice2011ArbitrarilyElements,Ewert20143Ancillae}
are all polarization independent, \ie\ polarization preserving with $U_h=U_v$, so this
restriction still allows to achieve non trivial discrimination.
The common polarization independence of these schemes was motivated by the symmetry of 
the Bell states set. We conjecture that the optimal interferometer presents the same symmetry, and is
polarization independent, except for an initial preprocessing step of the ancillæ, as for the single-photon schemes of \cite{Ewert20143Ancillae}.

We present this polarization preserving upper bound for a generic ancilla in Section \ref{sec:GeneUpB}, 
and bound it itself by a simple photon number dependent expression in Section~\ref{sec:photbnd}.

\subsection{Generic upper bound}
\label{sec:GeneUpB}

The ability of an interferometer to perform a BM is its ability to discriminate between the four Bell states
when they are supplied as its input, each with equal probability $1/4$.
We consider, \twlog, a pure $k$-photon additional
resource state $\ket{ϒ}=\sum_{λ=0}^{k} υ_λ \ket{ϒ,λ}$, 
where $\ket{ϒ,λ}$ is a pure state with $λ$ photons polarized horizontally 
and $k-λ$ photons polarized vertically.
Since our interferometer is polarization preserving, 
the total number of horizontally (or vertically) polarized photons is unchanged,
and this number can be easily deduced from the observed detection event.
Therefore, the measurement would be the same if one had performed a 
projective measurement of 
of these polarized photon numbers on the global input state, before feeding the projected state into the interferometer.

Let us rewrite the input state according to this projection:
\begin{align}
  \notag
  \ket{Ψ^±}\ket{ϒ}
    &=\sum_{λ=1}^{k+1}\frac{υ_{λ-1}}{\sqrt 2}(\ket{HV}±\ket{VH})\ket{ϒ,λ-1}\\ 
    \label{eqn:inprojphi}
  \ket{Φ^±}\ket{ϒ}
    &=\sum_{λ=0}^{k+2}\frac{υ_{λ-2}}{\sqrt 2}\ket{HH}\ket{ϒ,λ-2}
            ±\frac{υ_λ}{\sqrt 2}\ket{VV}\ket{ϒ,λ},
\end{align}
where we have set $υ_λ=0$ for $λ<0$ and $λ>k$, in order to include the edge 
cases in the formula.
Each term of the above sums corresponds to a term with $λ$ horizontally 
polarized photons. 

To obtain an upper bound, we assume the measurement to be perfect after the projection 
onto the state with $λ$ horizontally polarized photons and $(k+2)-λ$ vertically 
polarized ones, forgetting about the linear optics restriction. 
One can easily see in the above equations that, for each $λ$, 
the term corresponding to the states $\ket{Ψ⁺}$, $\ket{Ψ⁻}$, and $\ket{Φ^±}$
are in three orthogonal subspaces, with the only possible remaining ambiguity
being between the $\ket{Φ⁺}$ and $\ket{Φ⁻}$ states. 
Let us now look at the distinguishability of those states.

For an arbitrary ancillary state, the unambiguous discrimination has to 
be performed between the following  unnormalized states: 
\[ \ket{\Lambda^\pm} = \frac{υ_{λ-2}}{\sqrt 2}\ket{HH}\ket{ϒ,λ-2}±\frac{υ_λ}{\sqrt 2}\ket{VV}\ket{ϒ,λ}. \]
Unambiguous discrimination of such a pair of pure states is possible with
an optimal success probability~\cite{Ivanovic1987HowStates,Dieks1988OverlapStates,Peres1988HowStates} of 
$\norm{Λ}-\abs{\braket{\Lambda^+}{\Lambda^-}}$,  with $\norm{Λ}=\braket{Λ⁺}{Λ⁺}=\braket{Λ⁻}{Λ⁻}$.
This optimal success probability is an upper bound to what is achievable with linear optics and photon counting,
leading to
\begin{equation*} 
\Prb[\text{succ},λ]≤2\min\pars*{½\abs{υ_{λ-2}}²,½\abs{υ_λ}²}.
\end{equation*} 
The total success probability---assuming an equal input probability of $1/4$ for each Bell state,
and perfect discrimination of $\ket{\Psi^\pm}$---is then%
\begin{align*}
\PrbSu≤\tfrac12 + \tfrac12&∑_{λ=0}^{k+2}\min\pars*{\abs{υ_{λ-2}}²,\abs{υ_λ}²}\\\allowbreak
	=\tfrac12 + \tfrac12&∑_{\mathclap{λ\text{ even}}}\min\pars*{\abs{υ_{λ-2}}²,\abs{υ_λ}²}\\
     &+\tfrac12∑_{\mathclap{λ\text{ odd}}}\min\pars*{\abs{υ_{λ-2}}²,\abs{υ_λ}²}.
\end{align*}
Since the minimum is taken every two value of $λ$, even and odd indices have to be considered separately.
It is then easy to notice that, among even 
(resp. odd) values of $λ$, all values of $\abs{υ_λ}²$ are 
counted once except the local maxima, which are omitted, and the local minima, which are counted twice, 
leading to
\begin{align}
\notag
 \PrbSu≤1-&\tfrac12∑_{\mathclap{\substack{λ\text{ even}\\\abs{υ_λ}²\text{ loc max}}}}\abs{υ_λ}² 
    	+\tfrac12∑_{\mathclap{\substack{λ\text{ even}\\\abs{υ_λ}²\text{ loc min}}}}\abs{υ_λ}²\\
\label{eq:PsuccUPloc}
   -&\tfrac12∑_{\mathclap{\substack{λ\text{ odd}\\\abs{υ_λ}²\text{ loc max}}}}\abs{υ_λ}² 
    	+\tfrac12∑_{\mathclap{\substack{λ\text{ odd}\\\abs{υ_λ}²\text{ loc min}}}}\abs{υ_λ}². 
\end{align}
This bound can then be simplified to the following looser bound on the failure probability
\begin{equation}
  \label{eq:PfailLB}
  \Prb[\text{fail}]
    ≥\frac12 \pars*{\max_{λ\text{ even}}\pars*{\abs{υ_λ}²}
           +\max_{λ\text{ odd}}\pars*{\abs{υ_λ}²}},
\end{equation}
the latter being equivalent
to \eqref{eq:PsuccUPloc} iff $\abs{υ_λ}²$ has a single local maximum over even
values of $λ$ and a single local maximum over odd values of $λ$.

In Section~\ref{sec:results}, we will compute this bound for different ancillary states $\ket{ϒ}$. But for now, 
we can already use the above equations to compute a bound depending only on $\ket{ϒ}$'s photon number $k$.

\subsection{Photon-number based upper bound} \label{sec:photbnd}

Let us now work out a bound that is independent of the specific form of $\ket{\Upsilon}$. 
If the number of photons $k$ in the ancilla is odd,
there are $\frac{k+1}{2}$ possible odd values for $λ$ for which $υ_λ\neq 0$, and
as many even values. Eq.~\eqref{eq:PsuccUPloc} then leads to the bound 
\[\Prb[\begin{subarray}{l}\text{fail}\\k\text{ odd}\end{subarray}]
  ≥ \frac1{k+1},
\]
which can only be achieved by ancillary states where all even values of $λ$ are
equiprobable, and all odd values of $λ$ are equiprobable.
Similarly, if $k$ is even, there are $\frac{k}{2}$ possible odd values for $λ$ such that $υ_λ\neq 0$, and
$\frac{k}{2}+1$ even ones. This leads to the bound
\[\Prb[\begin{subarray}{l}\text{fail}\\k\text{ even}\end{subarray}]
  ≥ \frac1{k+2},
\]
which can only be achieved by ancillary states where all values of $λ$ for which $υ_λ\neq 0$ 
are even and equiprobable.

Both limits above can be expressed by the formula
\begin{equation} \label{eqn:photonbnd}
\Prb[\text{fail}]≥\frac{1}{\ceil{k+1}_{\mathrlap{\text{even}}}},
\end{equation}
where $\ceil{⋅}_{\text{even}}$ is the smallest even integer greater or equal to
its argument.
For the trivial case $k=0$, this result is a special case of the Calsamiglia--Lütkenhaus theorem 
\cite{Calsamiglia2001MaximumAnalyzer}. For $k=1$, we find that a single extra photon does not help, 
at least with a polarization preserving interferometer.

A non-trivial example of state achieving the limit for even $k$
is the $\{2^{N+1}-2\}$-photon ancillary state 
$\ket{ϒ₁}_{\text{G}} \cdots \allowbreak \ket{ϒ_N}_{\text{G}}$
defined by Grice in \cite{Grice2011ArbitrarilyElements}.
Note that, except for the 2-photon state $\ket{ϒ₁}_{\text{G}}=\ket{Φ⁺}$, the Grice scheme 
needs to use GHZ-like states~\cite{Greenberger1989GoingTheorem} of up to $2^N$ dual-rail qubits.
The $\ket{ϒ_n}_{\text{EvL}}$ states defined Ewert and van Loock in \cite{Ewert20143Ancillae}
can be written in the same form of the $\ket{ϒ_n}_{\text{G}}$ 
states of \cite{Grice2011ArbitrarilyElements} with respect to the distribution of horizontally polarized photons; 
however, at variance with them, in order to attain the same $\PrbSu$ two copies of each one are required.
The photon-number dependent upper bound~\eqref{eqn:photonbnd} is therefore not tight for them,
even if the generic bound \eqref{eq:PfailLB} is.
However, the schemes by Ewert and van Loock start by independently interfering each of the output of an initial 
50:50 beamsplitter with half of the ancillary state.
This additional restriction changes the photon dependent bound, which is then achieved by the Ewert--van Loock schemes.

\section{\label{sec:prog}Linear optical network optimizer}

We present here our linear optical network optimizer, a program looking for the optical network
maximizing $\PrbSu$ for a given ancillary state. 
We first restate our problem in terms of second quantization
in Section~\ref{sec:2ndQuantLinOpt} before detailing our approach, which can be divided into two parts.
As explained in Section~\ref{sec:symbolic}, 
for each ancilla we want to  analyze, and for each input Bell state, we generate a symbolic expression for the probability amplitudes of all output events in terms of $U$.
Those functions, along with their gradient with respect to the $U$ entries, are heuristically optimized in order to reduce the number of operations needed.
The optimized functions are then translated into a low-level language and compiled.
Then, as exposed in Section~\ref{sec:numopt},
a constrained numerical optimization using a nonlinear method is performed.
Due to the heavy non-smooth character of $\PrbSu$, 
we construct a meaningful figure of merit, function of the previously obtained probability amplitudes.

While this can seem at first glance an overkill \textit{brute-force} approach, both steps present important symmetries that we can exploit, gaining up to two orders of magnitude in computation time in some cases, and reducing function complexity.

\subsection{Polynomial representation of the network}
\label{sec:2ndQuantLinOpt}
We can represent the input (output) state to a $n$-modes linear optical network through a polynomial in the input (output) modes creation operators~\cite{Fock1932KonfigurationsraumQuantelung,*[{english translation in }]Faddeev2004V.A.Theory}
\begin{align*}
\ket{\psi_\text{in}} &= P_\text{in}(a^\dag_1,\dots,a^\dag_n)\ket{0}, \\
\ket{\psi_\text{out}} &= P_\text{out}(c^\dag_1,\dots,c^\dag_n)\ket{0}.
\end{align*}
The effect of the network can be represented by a unitary transformation $U := (u_{ij})$ connecting the input and the output modes:
\begin{equation}\label{eqn:intoout}
a^\dag_i = \sum_{j=1}^n u_{ij} \, c^\dag_j \, .
\end{equation}
Notice that this only implements a subset of all the transformations of the modes allowed by quantum
mechanics, namely photon interferences.

At variance with the previous section, we represent dual rail encoding with distinct spatial modes instead of orthogonal polarization modes~\cite{Kok2007LinearQubits}.
The Bell states are represented by:
\begin{align}\label{eqn:bellstates}
\ket{\Phi^\pm} &= \frac{1}{\sqrt{2}} \Big( a^\dag_1 a^\dag_3 \pm a^\dag_2 a^\dag_4 \Big) \ket{0}, \\
\label{eqn:bellstates2}
\ket{\Psi^\pm} &= \frac{1}{\sqrt{2}} \Big( a^\dag_1 a^\dag_4 \pm a^\dag_2 a^\dag_3 \Big) \ket{0}.
\end{align}
Let $ B_\beta (a^\dag_1,\dots,a^\dag_4) $, $\beta = 1,\dots,4 $ be the polynomial associated with the
above states; the generic input to the network is then represented by the polynomial
\begin{equation}\label{eqn:geninput}
(P_\text{in})_\beta = B_\beta (a^\dag_1,\dots,a^\dag_4) \, Q(a^\dag_5, \dots, a^\dag_n),
\end{equation}
where $Q$ describes the ancillary state $\ket{\Upsilon}$. 

At the output of the network, an array of polarizing beamsplitters followed 
by photon number resolving detectors (PNRD) carries out a measurement in the computational basis.
A \textit{detection event} is described by the number of photons detected in each output mode, 
\eg\  $ 1020 $ is the event in which three photons are detected, one in the first mode and two in the third mode. 
This event corresponds to the output state
\begin{equation}\label{eqn:ex1020}
\ket{1020}_{\text{out}}=\frac{c₁^{\dagger}c₃^{\dagger 2}}{√{1!0!2!0!}}\ket0=\frac{c₁^{\dagger}c₃^{\dagger 2}}{√{2}}\ket0,
\end{equation}
and its probability can therefore be computed form the the squared norm of the corresponding 
monomial coefficient in the polynomial \eqref{eqn:geninput}.

A Bell measurement is \textit{unambiguous}  when at least one output event occurs with nonzero probability 
for only one of the four input Bell states (plus ancilla);
events meeting this condition are said to be \textit{discriminating}.
By writing $ p_\beta^e $ for the square modulus of the amplitude associated with the detection event $e$ when the input is the Bell state $\beta$, the total probability of successful discrimination is:
\begin{align} \label{eqn:succprob}
  \PrbSu &= \frac{1}{4} \sum_{\mathclap{e,\tilde{β}∈S}} p_{\tilde{\beta}}^e&
  \text{with }& S=\{e,\tilde{β}: ∀β≠\tilde{β}, p_\beta^e=0 \}.
\end{align}

\subsection{Symbolic computation} \label{sec:symbolic}
The purpose of the method presented here is to provide the optimum-finding algorithm described in the next subsection with a fast, optimized function returning all the detection event probabilities, along with their gradients with respect to the entries of $U$, from which a figure of merit $f(U)$ will be constructed.
Working on this separately, instead of directly using sums of permanents of submatrices of $U$~\cite{Scheel2005ConditionalNetworks}, enables us to carefully analyze the problem and implement some analytical shortcuts that will ultimately speed up the search for optima.

We use SymPy~\cite{Meurer2017SymPy:Python},
an open-source symbolic computation library for Python.
The main procedure is as follows:
given an input polynomial in the form~\eqref{eqn:geninput}, we implement the transformation~\eqref{eqn:intoout} by direct substitution.
We then regroup the resulting multivariate polynomial in $ c^\dag_1,\dots,c^\dag_n $ and we extract the coefficient of each monomial;
they correspond to the amplitudes of all possible detection events, and they are complex functions of the entries of $U$.
A straightforward combinatorial argument shows that there are 
\begin{equation}\label{eqn:numdetev}
 N = \binom{n + k + 1}{k+2}
\end{equation}
possible detection events for each input state, where $k$ is the number of photons in the ancillary state---making $k+2$ the total number of photons entering the network.
In the following, we will always assume $n \geq k+4$.

As a simple example of what the algorithm does, consider a network with $n=4$ modes, with the Bell state $\ket{\Phi^+} = B_1 \ket{0}$ as input state and no ancillary state ($k = 0$), with
\begin{equation*}
B_1 = \frac{1}{√2}\pars{a^\dag_1 a^\dag_3 + a^\dag_2 a^\dag_4} .
\end{equation*}
Let $U$ be the unitary matrix representing the network, the output polynomial is obtained upon performing the substitution in eq.~\eqref{eqn:intoout}:
\begin{equation}\label{eqn:exPout}
\begin{split}
P_\text{out} = &\frac{1}{√2}
				\Big( \sum_{j_1} u_{1j_1} c^\dag_{j_1} \Big)
				\Big( \sum_{j_2} u_{3j_2} c^\dag_{j_2} \Big) \\
             &+ \frac{1}{√2}
             	\Big( \sum_{j_3} u_{2j_3} c^\dag_{j_3} \Big)
                \Big( \sum_{j_4} u_{4j_4} c^\dag_{j_4} \Big).
\end{split}
\end{equation}
After expanding all the products, we obtain a polynomial in $c^\dag_1,\dots,c^\dag_4$ with $N=10$ terms of degree $k+2=2$.
The coefficient of the monomial $c^\dag_1 c^\dag_3$ is, for example, the amplitude of the detection event 1010.
For an arbitrary event, with $k_i$ photons in mode $i$, a bosonic correction factor $\sqrt{\prod_i k_i!}$ has to be applied,
as in eq.~\eqref{eqn:ex1020}.
Expanding~\eqref{eqn:exPout}, all the output event amplitudes read:
\begin{align*}
& 2000 && \longrightarrow && u_{11}u_{31} + u_{21}u_{41} \\
& 0200 && \longrightarrow && u_{12}u_{32} + u_{22}u_{42} \\
& 0020 && \longrightarrow && u_{13}u_{33} + u_{23}u_{43} \\
& 0002 && \longrightarrow && u_{14}u_{34} + u_{24}u_{44} \\
& 1100 && \longrightarrow && (u_{11}u_{32} + u_{12}u_{31} + u_{21}u_{42} + u_{22}u_{41})/√2 \\
& 1010 && \longrightarrow && (u_{11}u_{33} + u_{13}u_{31} + u_{21}u_{43} + u_{23}u_{41})/√2 \\
& 1001 && \longrightarrow && (u_{11}u_{34} + u_{14}u_{31} + u_{21}u_{44} + u_{24}u_{41})/√2 \\
& 0110 && \longrightarrow && (u_{12}u_{33} + u_{13}u_{32} + u_{22}u_{43} + u_{23}u_{42})/√2 \\
& 0101 && \longrightarrow && (u_{12}u_{34} + u_{14}u_{32} + u_{22}u_{44} + u_{24}u_{42})/√2 \\
& 0011 && \longrightarrow && (u_{13}u_{34} + u_{14}u_{33} + u_{23}u_{44} + u_{24}u_{43})/√2 \, .
\end{align*}
The probabilities $p_\beta^e$ are then obtained by taking the square moduli of those amplitudes.
All the above steps are automated in our program, by exploiting polynomial manipulation routines contained in SymPy; we just have to `manually' feed as input the ancillary polynomial $Q$.

The numerical optimization routine also needs the Jacobian of the
figure of merit, and the above approach allows us to compute it symbolically in order to 
speed up the optimization. This computation is done through the symbolic derivatives 
of $p_\beta^e$ with respect to the real and the imaginary part of $u_{ij}$.
The $p_\beta^e$ are the square modulus of a complex holomorphic function (more precisely, a polynomial) and, for any such function $g(u_{11},\dots,u_{nn})$, we have
\begin{equation*}
\pdv{\abs{g}^2}{\Re{u_{ij}}} = 2 \Re{g \pdv{g^*}{u_{ij}}}
\end{equation*}
and an analogous expression for the derivative with respect to the imaginary part.
This allows for a simple symbolic computation of the relevant derivatives.
If the optimization algorithm were not provided with a Jacobian function, it would have had to estimate it by evaluating the figure of merit at nearby points. 
Using the finite differences method, this amounts to at least $2n^2$ additional evaluations of $f(U)$ per iteration---to be compared with evaluating $n^2$ symbolic derivatives, each of which is a strictly simpler function (\ie{} a polynomial with less terms) than the corresponding probability amplitude.
Furthermore the Jacobian would only be calculated to some fixed accuracy, which adds noise to the optimization algorithm and worsen its convergence.

As the output of this computation, we need $4N$ expressions for the probabilities, each one with $n^2$ expressions for the derivatives.
But there is no need to actually compute the symbolic form of all the $ 4N $ events for each ancilla.
Two main simplifications help to drastically reduce the number of symbolic computations:
\begin{enumerate}
\item
  The detection events divide into equivalence classes under permutation of the output modes, 
  corresponding to a permutation of the columns of $U$.
  It is easy to check that, of all the events in the above example, it is only necessary to obtain 2000 and
  1100. For example, 1010 can be obtained from 1100 by swapping the second and the third column of $U$, at a negligible computational cost.
  The number of independent events does not depend on $n$ and is equal to $P_{k+2}$, the number of integer 
  partitions\footnote{\emph{I.e.\@} the number of positive integer sums equal to $k+2$.}~\cite{Abramowitz1988HandbookTables} 
  of the total number of photons.
  This also reduces the number of gradients to calculate, from $n^2$ to at most $n(k+2)$.
\item 
  As it is clear from eqs.~\eqref{eqn:bellstates} and \eqref{eqn:bellstates2} , knowing 
  an event's amplitudes for a single Bell state allows one to
  compute its amplitudes for all of them, just by swapping two rows of $U$ and/or changing their sign.
  This reduces the number of functions to compute by a further factor of $4$.
\end{enumerate}
With these expedients, we reduce the problem to the computation of just $P_{k+2}$ probability functions 
along with their gradients.
For example, when $n=8$, $k=2$ as in the first scheme of Grice's paper (using a $\ket{Φ⁺}$ ancillary state), 
we end up decreasing the number of function to compute from $ 4N = 1320 $ to only $ P_4 = 5 $, 
each with at most 32 derivatives instead of 64.

As SymPy is written in pure Python, we can accelerate this section of the program using PyPy~\cite{PedroniGoalsDocumentation}, an implementation of the Python interpreter with a Just-In-Time compiler.
For large networks ($n\geq 8, k\geq 2$) we obtain a tenfold speedup over plain Python, at the cost of increased memory usage%
\footnote{As an example, first iteration of the Ewert--van Loock scheme
($\ket{1}^{⊗4}=\ket{ϒ₁}_{\text{EvL}}^{⊗2}$) takes 7 minutes and 30 seconds on our laptop (see Table~\ref{tab:spec})
using the standard Python interpreter and about 150 MB of RAM.
Using PyPy the time is cut down to 45 seconds, with a memory consumption of 250 MB.}.
Even with all these optimizations in place, however, the problem still scales exponentially%
\footnote{We nonetheless get an exponential advantage over the naive approach, as we can see from the asymptotic formula for $P_k$ due to Hardy and Ramanujan~\cite{Hardy1918AsymptoticAnalysis}:
\[ P_k \underset{k \rightarrow \infty}{\sim} \frac{1}{4k\sqrt{3}}\exp{\pi\sqrt{\frac{2k}{3}}}, \]
to be compared with the binomial coefficient in eq.~\eqref{eqn:numdetev}, lower bounded by $2^{k+2}$.}.
For comparison, we obtain the functions for the first Grice iteration (``one extra Bell pair'' in Table~\ref{tab:results}) in about 6 seconds using 100 MB of RAM;
the second iteration of Grice's scheme---the largest calculation we managed to complete,
with $n=16$ and $k=6$---took instead 10 days of single-core CPU time on the cluster described in Table~\ref{tab:spec}, requiring a large portion of the 256 GB of RAM at our disposal.
The resulting (already heavily optimized) probability functions for this case consist of a grand total of about $1.8$ \textit{million} addition and multiplication operations, and the Jacobian of about $17$ million.

We use Theano~\cite{TheanoDevelopmentTeam2016Theano:Expressions}, a numerical computation library for Python, 
as our backend in order to translate each function into C and compile it to a fast numerical version.

\subsection{Numerical optimization}
\label{sec:numopt}
\begin{table}[tb]
\renewcommand\cellgape{\Gape[2pt]}
\caption{Specifications of the two computers used in this article. 
  The frequency is the nominal frequency of the processor. 
  The cluster is the \texttt{gmpcs-206} branch
  of the computing center MésoLUM of the LUMAT research federation~\cite{MesocentreMesoLUM},
  and the specifications refer to a single node.}
\label{tab:spec}
\begin{ruledtabular}
\begin{tabular}{ccccc}
Name & Processor & \# of & Freq.\@  & RAM\\ 
     &  model    & cores & (GHz) &    (GB)\\ 
\hline
Laptop & \makecell{Intel Core \\ i7-4710MQ} & 4 & 2.5 & 16 \\
Cluster & \makecell{Intel Xeon \\  E5-2670}  & 12 & 2.3 & 256
\end{tabular}
\end{ruledtabular}
\end{table}

The second part of the procedure takes care of finding the optimal value of $\PrbSu$ in eq.~\eqref{eqn:succprob} for each input ancillary state.
Naively, we could straightforwardly calculate $\PrbSu(U)$ from the numerical evaluation of the probability functions obtained in the previous section;
however, $\PrbSu(U) = 0$ almost everywhere in the domain $U(n)$, and it is neither continuous nor differentiable in the 
region of interest, where $\PrbSu(U) ≠ 0$.
This obviously makes most optimization methods highly ineffective.

As a workaround, we devise a figure of merit as a continuous alternative to $\PrbSu(U)$.
We thus search for local \textit{minima} of:
\begin{equation}\label{eqn:fom}
f(U) = \sum_e \Big( \sum_\beta p_\beta^e - 2 \max_\alpha p_\alpha^e \Big),
\end{equation}
of which the addends of the outer sum are equal to the ones of $-\PrbSu(U)$, when the latter happen to be nonzero.
Ideally, we want the figure of merit to closely mimic the behavior of $\PrbSu$ around the optima. In particular it would be useful to have the following holding for each pair of locally optimal unitaries $U_1$ and $U_2$:
\begin{align*}
f(U₁) < f(U₂) && \text{iff} && \PrbSu(U₁) > \PrbSu(U₂).
\end{align*}
Unfortunately, the mutual relation between $f(U)$ and $\PrbSu(U)$ is not simple.
While it seems reasonable to conjecture the set of local minima of $f(U)$ to include the set of $\PrbSu(U)$'s maxima, we find that, for some $U_1$ and $U_2$:
\begin{align} \label{eqn:fomrel}
f(U₁) < f(U₂) && \text{while} && \PrbSu(U₁) < \PrbSu(U₂).
\end{align}
This forces us to conduct a search among all local optima of $f(U)$, rather than taking advantage of global optimization methods like simulated annealing.

The optimization itself is implemented using SciPy’s \textit{Sequential Least-Square Programming} (SLSQP) optimization method~\cite{Oliphant2007SciPy:Python,Kraft1988AProgramming}.
As this method supports equality constraints, we represent $U$ through $ 2n^2 $ real variables describing its entries' real and imaginary parts, along with $n^2$ equations enforcing orthonormality of the rows.
In order to improve convergence speed, the method is allowed some leniency on the constraints, in that they only have to be satisfied at the local optimum.
In order to ensure uniform sampling of the starting points of each optimization, we choose them by picking a random matrix from the Haar measure on the unitary group $U(n)$.
This is accomplished using the QR decomposition method described in \cite{Ozols2009HowMatrix}.

We compared the performance of the constrained method above to the  Broyden-Fletcher-Goldfarb-Shanno (BFGS) 
algorithm~\cite{Nocedal2006NumericalOptimization}, a popular unconstrained quasi-Newton method.
While the latter uses no constraints and has to work with independent variables%
\footnote{We encoded the $n^2$ unconstrained degrees of freedom of $U(n)$ into an Hermitian matrix $H$, using $U(n)=e^{iH}$.}%
, the relation between those and the entries of $U$ is non-trivial and this therefore hinders our ability to input the analytical form of the gradient.
Thus the performances are comparable with SLSQP, if not worse in some cases, even if the number of variables is cut by half.
Furthermore, the convergence accuracy and precision seems unaffected by the choice of one method over the other.

\section{Results}
\label{sec:results}

\begin{table*}[tbh]
\vspace*{-2\baselineskip}
\caption{\label{tab:results} Summary of known analytical and numerical results for different ancillæ. 
$Q$ is the input polynomial, $n$ the number of modes and $k$ the number of photons.
$\PrbSu^{\text{num}}$ is the optimum obtained through our optical network optimizer; 
the fraction given is exact up to our numerical precision (9 decimals).
$\PrbSu^{\text{ana}}$ is the best known explicit analytical result.
$\PrbSu^{\text{upp}}$ and $\PrbSu^{\text{upp}}(k)$ are our analytical upper bounds for polarization-preserving 
networks (section~\ref{sec:UppBPolInd}), for the ancilla and for arbitrary ancillæ with same $k$.
They are in bold font when matching the best known result.}

\setlength{\tabcolsep}{3.9pt}
\renewcommand{\arraystretch}{1.85}
\renewcommand\theadfont{\bfseries}

\begin{tabular}{lccccccc} \hline\hline \\[-22pt]
\thead{State}		& $\bm Q$
		&\thead{$ \bm n $} & \thead{$ \bm k $}
		& \thead{$\PrbSu^{\text{num}}$}
		& \thead{$\PrbSu^{\text{ana}}$}
		& \thead{$\PrbSu^{\text{upp}}$}
        & \thead{$\PrbSu^{\text{upp}}(\bm{k})$}\\[-1pt]
\hline
\multicolumn{4}{l}{\textbf{Vacuum Ancilla}}\\[-4pt]
$\ket{0}$&$1$
		&4--14 & 0 
        & 1/2
		& 1/2\cite{Calsamiglia2001MaximumAnalyzer}
        & \textbf{1/2}\footnote{Also holds for polarization non-preserving interferometers.}\cite{Calsamiglia2001MaximumAnalyzer}
        & \textbf{1/2} \\ 
\hline 
\multicolumn{4}{l}{\textbf{$\bm{k/2}$ Extra Bell Pairs}} \\[-4pt]
$\ket{Φ⁺}^{⊗k/2}$        & $ \frac{(a^\dag_5 a^\dag_7 + a^\dag_6 a^\dag_8) \dots
			(a^\dag_{2k+1} a^\dag_{2k+3} + a^\dag_{2k+2} a^\dag_{2k+4})}{2^{k/4}} $ 
&$2k+4$ & even
        & ---
        & \footnote{\label{foot:unknown}No generic scheme is known.}
        & \makecell{ \\[-15pt] $1-\frac{\binom{k/2}{\floor*{k/4}}}{2^{ k/2 +1 }}$
          \footnote{\label{foot:pifourth}Polarization-preserving bound obtained after rotating 
        	  the polarization of some or all modes by $\frac{\pi}{4}$.}
        \\[2pt] $\simeq 1-\tfrac{1}{\sqrt{πk}}$}
        & $\displaystyle\frac{k+1}{k+2}$\\
$\ket{Φ⁺}=\mathrlap{\ket{ϒ₁}_{\mathrm{G}}}$ &$\frac{(a^\dag_5 a^\dag_7 + a^\dag_6 a^\dag_8)}{\sqrt{2}} $
&8 & 2
        & 3/4
        & 3/4\cite{Grice2011ArbitrarilyElements}
		& \textbf{3/4}
        & \textbf{3/4} \\ 
$\ket{Φ⁺}^{⊗2}$		& $\frac{(a^\dag_5 a^\dag_7 + a^\dag_6 a^\dag_8) (a^\dag_9 a^\dag_{11} + a^\dag_{10} a^\dag_{12})}{2} $ 
&12 & 4
		& 3/4
        & \footnote{\label{foot:same}The best known interferometer correspond to a smaller ancilla, together with ignoring extra modes.}
        & \textbf{3/4} 
        & 5/6 \\
$\ket{Φ⁺}^{⊗3}$		& $ \frac{(a^\dag_5 a^\dag_7 + a^\dag_6 a^\dag_8) \dots
			(a^\dag_{13} a^\dag_{15} + a^\dag_{14} a^\dag_{16})}{\sqrt{8}} $ 
&16 & 6
		& \footnote{\label{foot:heavy}Computation out of reach for our program.}
        & \footref{foot:same}
        &  13/16
        &  7/8 \\ 
\hline 
\multicolumn{3}{l}{\textbf{$\bm k$ Extra Photons}} \\[-4pt]
$\ket{1}^{⊗k}$        &$ a^\dag_5 \dots a^\dag_{k+4} $
&$k+4$ & even
        & ---
        & \footref{foot:unknown}
        & \makecell{ \\[-15pt] $1-\frac{\binom{k/2}{\floor*{k/4}}}{2^{ k/2 +1 }}$
          \footref{foot:pifourth}
        \\[2pt] $\simeq 1-\tfrac{1}{\sqrt{πk}}$}
        & $\displaystyle\frac{k+1}{k+2}$
\\
$\ket{1}^{⊗k}$        &$ a^\dag_5 \dots a^\dag_{k+4} $
&$k+4$ & odd
        & ---
        & \footref{foot:unknown}
        &\multicolumn{2}{c}{\footnotesize{same as above, for $k-1$}}
        \\
$\ket{1}$		& $ a^\dag_5 $ 
&5 & 1
		& 1/2\footref{foot:same}
        & \footref{foot:same}
        & \textbf{1/2}
        & \textbf{1/2} \\ 
$\ket{1}^{⊗2}$		& $ a^\dag_5 a^\dag_6 $
&6 & 2
        & 5/8
		& 5/8
        &  3/4\footref{foot:pifourth} 
           (\textbf{5/8})\footnote{\label{foot:sym}For networks which start by interfering the two bell states on a 
            50:50 beamsplitter, analyzing each half separately.}
        & 3/4\\ 
$\ket{1}^{⊗3}$		& $ a^\dag_5 a^\dag_6 a^\dag_7$ 
&7 & 3
        & 5/8
		& \footref{foot:same}
        & 3/4\footref{foot:pifourth} 
            (\textbf{5/8})\footref{foot:sym}
        & 3/4\\ 
$\ket{1}^{⊗4}\mathrlap{=\ket{ϒ₁}^{⊗2}_{\mathrm{EvL}}}$		& $ a^\dag_5 a^\dag_6 a^\dag_7 a^\dag_8 $ 
&8 & 4
        &3/4
		& 3/4\cite{Ewert20143Ancillae}
		& \textbf{3/4}\footref{foot:pifourth}
        & 5/6\\ 
$\ket{1}^{⊗6}$
& $ a^\dag_5 \dots a^\dag_{10} $ 
&10 & 6
        & 3/4
		& \footref{foot:same}
        & 13/16\footref{foot:pifourth}
        & 7/8\\
$\ket{1}^{⊗8}$		& $ a^\dag_5 \dots a^\dag_{12} $ 
&12 & 8
		& \footref{foot:heavy}
        &49/64
        &  13/16\footref{foot:pifourth} \!\!\! (25/32)\footref{foot:sym}
        & 9/10\\ 
$\ket{1}^{⊗12}$		& $ a^\dag_5 \dots a^\dag_{16}$ 
&16 & 12
        &\footref{foot:heavy}
		& 25/32\cite{Ewert20143Ancillae} 
        & 27/32\footref{foot:pifourth} \!\!\! (13/16)\footref{foot:sym}
        & 13/14\\

\hline 
\multicolumn{4}{l}{\textbf{Grice Schemes \cite{Grice2011ArbitrarilyElements}} (first iteration is $\ket{\Phi^+} = \ket{\Upsilon_1}_G$ above)}\\[-2.5pt]
$\ket{ϒ₁}_{\mathrm{G}}⋯\ket{ϒ_N}_{\mathrm{G}}$
& {\footnotesize \, Straightforward, but long expression}
& $2k+4$ & $2^{N+1}\!\!-2$
        & ---
        & $\displaystyle\frac{k+1}{k+2}$
        & $\bm{\displaystyle\frac{k+1}{k+2}}$ 
        & $\bm{\displaystyle\frac{k+1}{k+2}}$ \\ 
$\ket{ϒ₁}_{\mathrm{G}}\ket{ϒ₂}_{\mathrm{G}}$
&$\frac{(a^\dag_5 a^\dag_7 + a^\dag_6 a^\dag_8)
			(a^\dag_9 a^\dag_{11} a^\dag_{13} a^\dag_{15} + a^\dag_{10} a^\dag_{12} a^\dag_{14}a^\dag_{16})}{2} $ 
&16 & 6
        & 9/16\footnote{Computation at the borderline of our computing capacity: this result is the best of just 12 optimizations over the course of three weeks.}%
        \footnote{\label{foot:worsethan}Numerical result worse than the best known analytical scheme.}
		& 7/8
        & \textbf{7/8} 
        & \textbf{7/8} \\
\hline 
\multicolumn{4}{l}{\textbf{Ewert--van Loock Schemes \cite{Ewert20143Ancillae}} (first iteration is $\ket{1}^{⊗4}=\ket{ϒ₁}^{⊗2}_{\mathrm{EvL}}$ above)}\\[-2.5pt]
\!\!\!$(\ket{ϒ₁}_{\mathrm{EvL}}\!\cdots\! \ket{ϒ_N}_{\mathrm{EvL}})\mathrlap{^{\otimes 2}}$
& {\footnotesize \, Straightforward, but long expression}
& $k+4$ & $2^{N+2}\!\!-4$
        & ---
        & $\displaystyle\frac{k+2}{k+4}$
        & $\displaystyle\bm{\frac{k+2}{k+4}}$
        & \hspace*{-17pt} $\displaystyle\frac{k+1}{k+2} \!
        		\left(\bm{\frac{k+2}{k+4}}\right)\!\!\raisebox{5pt}{\footref{foot:sym}}$ \hspace*{-10pt}  \\[4pt]
\hline 
\multicolumn{4}{l}{\textbf{GHZ states}}\\[-4pt]
$\ket{\mathrm{GHZ}_k}$		& $\frac{a^\dag_5 \cdots a^\dag_{2k+3}
        				+ a^\dag_6 \cdots a^\dag_{2k+4}}{\sqrt{2}}$ 
&$2k+4$ & $k$
		& ---
        & 3/4\footnote{\label{foot:backtoBell}Success probability achieved by measuring all photons of the 
        	ancilla and using the remaining in a ``one extra Bell pair'' scheme.}
		& \textbf{3/4}\footref{foot:pifourth}
        & $\hspace*{-9pt} 1\!-\!\frac{1}{\ceil{k+1}_{\mathrlap{\text{even}}}}$ \\ 
$\ket{\mathrm{GHZ}_3}$ 		& $ \frac{a^\dag_5 a^\dag_7 a^\dag_9 + a^\dag_6 a^\dag_8 a^\dag_{10}}{\sqrt{2}} $ 
&10 & 3
		& 3/4
        & 3/4\footref{foot:backtoBell}
		& \textbf{3/4}\footref{foot:pifourth}
        & 3/4 \\
$\ket{\mathrm{GHZ}_4}\mathrlap{=\ket{ϒ₂}_{\mathrm{G}}}$		& $\frac{a^\dag_5 a^\dag_7 a^\dag_9 a^\dag_{11}
        				+ a^\dag_6 a^\dag_8 a^\dag_{10} a^\dag_{12}}{\sqrt{2}}$ 
&12 & 4
		& 3/4
        & 3/4\footref{foot:backtoBell}
		& \textbf{3/4}\footref{foot:pifourth}
        & 5/6 \\ 
\hline 
\multicolumn{4}{l}{\textbf{W State}}\\[-4pt]
$\ket{\mathrm{W₃}}$ 
	& $\frac{a^\dag_6 a^\dag_7 a^\dag_9
						+ a^\dag_5 a^\dag_8 a^\dag_9 + a^\dag_5 a^\dag_7 a^\dag_{10}}{\sqrt{3}}$
	&\makecell{10--11 \\[\jot] 12--14 }
    & 3
    &\makecell{5/9\footref{foot:worsethan} \\[\jot]
    \scalebox{0.95}{$\mathclap{0.5785508(2)\footref{foot:worsethan}}$} }
    & 7/12
    & {2/3}\footref{foot:pifourth} (3/4)\footnote{Obtained through a more complex transformation of the input, exposed in the main text.}
    & 3/4
        \\[2\jot] \hline\hline
\end{tabular}
\end{table*}

Below, we work out the analytical upper bound for polarization-preserving interferometers in Section~\ref{sec:UppBPolInd} for the states we used, and we compare them to the numerical results we obtained for generic interferometers.

Some of the optimization results for different input ancillæ 
$\ket{\Upsilon} = Q(a^\dag_5,\dots,a^\dag_n) \ket{0}$ are summarized in Table~\ref{tab:results}.
For each one of them, we collected the local optima from about ten thousand successful iterations of the optimization algorithm.
In the table the maximum value achieved for each input is shown;
it can be noted that, for the cases already known in the literature, we find the same maximal discrimination probability.
Furthermore, we were able to work with other types of ancillæ.

\subsection{Vacuum extra modes} \label{sec:vacuumextra}
By virtue of the Calsamiglia–Lütkenhaus theorem \cite{Calsamiglia2001MaximumAnalyzer}, the analytical upper bound of $\PrbSu \leq 1/2$ is known to hold for general interferometers, equipped with an ancilla consisting of an unlimited number of extra modes in the vacuum state.
We can work out different version of this bound (for photon-number- and polarization-preserving measurements) following the reasoning laid out in Section~\ref{sec:UppBPolInd}, looking at the distinguishability of the states in eq.~\eqref{eqn:inprojphi} after a projection onto the basis of the number-of-horizontally-polarized-photons operator.
If the ancillary state is the vacuum,
or any other state with a fixed number $\overline{λ}$ of horizontally polarized photons,
$υ_{\overline{λ}}$ is the only value of $λ$ for which $υ_{λ}\neq 0$.
This leads to just two distinct terms: 
\begin{align}
\label{eqn:vacfirst} &\ket{HH}\ket{ϒ} && \text{for both $\ket{Φ⁺}$ and $\ket{Φ⁻}$}, \\
\label{eqn:vacsec} ±&\ket{VV}\ket{ϒ} && \text{respectively for $\ket{Φ⁺}$, $\ket{Φ⁻}$}.
\end{align}
The term~\eqref{eqn:vacfirst} is identical for both inputs, while the~\eqref{eqn:vacsec} differs by a global phase.
Thus, these terms are not distinguishable at all: in this case
the ancillary state cannot help to discriminate between the two different inputs,
and $\PrbSu≤ 1/2$.

Indeed, without extra photons the maximum discrimination probability that we find through our numerical optimization is~$1/2$, for any value of $n$ we tried.
We quickly achieve this maximum on our laptop (see Table~\ref{tab:spec}), and we collect a thousand successful iterations in a matter of minutes for different values of $n≤14$.
The $n=14$ case still takes less than an hour on the laptop, and a few minutes of the cluster.

\subsection{Extra Bell pairs} \label{sec:extraBP}
One of the simplest linear optical network schemes with ancilla achieving more-than-$\tfrac12$ Bell state discrimination probability is arguably the first iteration of Grice's strategy \cite{Grice2011ArbitrarilyElements}.
He shows that adding one extra $\ket{\Phi⁺}$ as ancilla helps cutting the degeneracy of $\ket{\Phi^\pm}$ by half, achieving $\PrbSu=3/4$.
However, the states used by Grice to increase its success probability past $3/4$ become more complex at each iteration,
since each additional ancillary state $\ket{ϒ_N}_\text{G}$ is a $2^N$-photon GHZ state. 
It would be experimentally much simpler to use multiple Bell pairs $\ket{Φ⁺}^{⊗k/2}$ as a $k$-photon ancilla,
which motivate our research of schemes using such resources.

We start by working out the polarization-preserving bound of Section~\ref{sec:UppBPolInd}.
We restrict ourselves to a product of $k/2$ states of the form $\ket{ϒ₁} = \frac{1}{\sqrt{2}} (\ket{2H}+\ket{2V}) $, where $\ket{2H}$ (resp.\@ $\ket{2V}$) is any state of two horizontally (resp.\@ vertically) polarized photons.
Bell pairs are a special case of the latter, as well as the $\ket{\Upsilon_1}_\text{EvL}$ defined in~\cite{Ewert20143Ancillae}.
We have
\begin{align*}
  \ket{ϒ₁}^{⊗k/2}
    &=2^{-k/4} (\ket{2H}+\ket{2V})^{⊗k/2}\\
    &=2^{-k/4}\sum_{λ=0}^{k/2}\sqrt{\binom{k/2}{λ}}\ket{ϒ, 2λ},
\end{align*}
where $\ket{ϒ, 2λ}$ is the uniform superposition of the terms with $2λ$ 
horizontally polarized photons present in the expansion of~$\ket{ϒ₁}^{⊗k/2}$.
Equation \eqref{eq:PsuccUPloc} then only have even nonzero terms, and this leads to%
\begin{align}
  \Prb[\begin{subarray}{l}\text{fail}\\\ket{Φ⁺}^{⊗k/2}\end{subarray}]
    &\geq 2^{- k/2 -1 }\binom{k/2}{\floor*{k/4}} \label{eq:Psep} \\
    &\geq 2^{- k/2 -1 }\frac{1}{\sqrt{\pi k}} \, 2^{\,k/2+1} \e^{-2/3k} \notag \\
    &=\frac{1}{\sqrt{\pi k}} \, \e^{-2/3k} \label{eqn:Pstir}, 
\end{align}
where we have supposed $k$ to be a multiple of $4$ and applied a second-order version of Stirling’s approximation~\cite{Robbins1955AFormula}
\begin{equation*}
\sqrt{2\pi n} \pars*{\frac{n}{\e}}^n
\leq n! \leq
\sqrt{2\pi n} \pars*{\frac{n}{\e}}^n \e^{1/12n}
\end{equation*}
When $k$ is even, but not a multiple of $4$, 
the inequality~\eqref{eqn:Pstir} is invalid, but we still have
\begin{align}
\Prb[\begin{subarray}{l}\text{fail}\\\ket{Φ⁺}^{⊗k/2}\end{subarray}] \geq
\frac{1}{\sqrt{\pi k}}\pars*{1+O\pars*{\frac{1}{k}}\!}.
\end{align}
The $1/\sqrt{k}$ scaling of $\Prb[\text{fail}]$ allowed by the above bound is worse than the $1/k$ scaling achieved by Grice schemes.
Nevertheless, it does not rule out strategies approaching success probabilities arbitrarily close to 1 by using as inputs much simpler states, \ie\ $k/2$ Bell pairs.

Our numerical search, which is not restricted to polarization-preserving schemes, 
converges in just about a minute to the $\PrbSu = 3/4$ scheme on our laptop, using 300~MB of RAM.
Unfortunately, the use of two extra Bell pairs shows no improvement over a single extra Bell pair, and its optimization uses
significantly more resources: about 4 hours with 20 parallel threads on the cluster (see Table~\ref{tab:spec}), 
each using 3~GB of RAM, for the collection of a thousand optimizations.
For three extra-bell pairs, the polarization-preserving bound of eq.~\eqref{eq:Psep} gives $\PrbSu≤13/16=.8125$ for the latter, allowing in principle for a scheme beyond $3/4$.
However, this dimensionality is barely out of reach for our program, even using the cluster. For comparison with a similar-sized case, the second iteration of Grice's strategy (the symbolic function's sheer size of which we discussed at the end of Section~\ref{sec:symbolic}) takes about 48 hours on the cluster for each starting point to converge to a local optimum.
We collected just 12 optimizations, obtaining $\PrbSu=9/16$; unfortunately this result is well below the known Grice's 7/8 scheme.

\subsection{Extra single photons}

The possibility of improving the discrimination probability through the use of unentangled extra single photons
is of great experimental interest, especially with
the recent development of high-efficiency single photon sources with near ideal 
indistinguishability~\cite{Senellart2017High-performanceSources}: 
such ancillary states would
be among the simplest types of input states for a real-world implementation of
linear optical Bell measurements.

Ewert and van Loock explore the use of pairs of single photon per auxiliary dual-rail mode~\cite[Section~D of supp.\@ mat.\@]{Ewert20143Ancillae} as substitutes of their ancillary state $\ket{\Upsilon_1}_\text{EvL}$.
While the initial transformation they apply to the input photons is polarization dependent, we can still use the formalism of our polarization-preserving upper bound of Section~\ref{sec:UppBPolInd}, restricted to the case in which each photon enters the network polarized along the $±\frac{π}4$ direction.
In the horizontal-vertical basis, this ancilla is described by the Hong-Ou-Mandel state~\cite{Hong1987MeasurementInterference}
$\ket{ϒ_1}_{\text{EvL}}=\frac1{√2}(\ket{20}+\ket{02})$
in each mode pair.
The $υ_λ$ coefficients in the case of the $k$-photon state (with $k$ even) $\ket{ϒ_1}_{\text{EvL}}^{⊗k/2}$
correspond to the ones of $k/2$ Bell states $\ket{Φ⁺}^{⊗k/2}$.
We can therefore apply the same reasoning laid out in Section~\ref{sec:extraBP}, obtaining the bound in eq.~\eqref{eq:Psep}.

With this restriction in place, we get for $k\geq 4$ a slightly tighter lower bound to $\Prb[\text{fail}]$, compared to the photon-number based bound~\eqref{eqn:photonbnd}.
For example, with 4 single photons the latter gives $\Prb[\text{fail}] \geq 1/6$, while eq.~\eqref{eq:Psep} gives $\Prb[\text{fail}] \geq 2^{-3} \binom{2}{1} = 1/4$.
In fact, the bound is saturated by the 4-single-photon variant of the first iteration of Ewert-van Loock strategy.
They also consider the 12-single-photon state $\ket{1}^{⊗12}→\ket{ϒ₁}^{⊗6}$.
In this case, a direct application of equation \eqref{eq:Psep} leads to $\Prb[\text{fail}]≥2^{-7}\binom{6}{3}=5/32$,
which is indeed smaller than the actual $7/32$ failure rate found by the authors.
However a look into the detailed symmetry of the strategy, as per the same reasoning used at the end of Section~\ref{sec:photbnd}, leads to the better (but more restricted) bound $\Prb[\text{fail}]≥2^{-4}\binom{3}{1}= 3/16$, which is closer, but still below $7/32$.

The aforementioned 4-photon scheme discriminates $\ket{\Psi^+}$ and $\ket{\Psi^-}$ with certainty, and $\ket{\Phi^\pm}$ only half of the times.
Our numerical algorithm indeed finds this $(1, 1, \tfrac12, \tfrac12)$ scheme when initialized with a 4 single photon ancilla, and does not manage to improve its probability of success---an evidence of its optimality even in the polarization-dependent case.
Furthermore, we find other schemes achieving the same total success probability with a different discrimination pattern among the Bell states: $(1,\tfrac34,\tfrac34,\tfrac12)$.

\begin{figure}[tb]

\caption{The first ``half'' Ewert–van Loock single-photons scheme.
It performs a Bell measurement with $\PrbSu=5/8$ on the state $\ket{\beta}$ using two unentangled extra photons $\ket{1}$, two polarization-independent beamsplitters, four polarizing beamsplitters, two phase shifters and six photocounters.\\}
\label{fig:halfewvan}

\begin{tikzpicture}[use optics,rotate=90,transform shape]
\tikzstyle{BS} = [thick optics element,rotate=90,scale=0.5,xshift=-3]
\tikzstyle{PS} = [thick optics element,rotate=-45,scale=0.3,object aspect ratio=0.3,label={[shift={(-0.10,0.2)},rotate=-90] $\sfrac{\lambda}{2} $ }]
\tikzstyle{PBS} = [beam splitter,rotate=-45,scale=0.62]

\tikzset{ PNRD/.pic= {
	\node (-c) at (0,0) {};
	\filldraw[fill=gray] (-0.5,0) -- (0.5,0) arc(0:180:0.5) -- cycle ;
	\draw[decorate,decoration=snake] (0,0.5) -- ++(0,0.5);
    }
}

\tikzset{ PRD/.pic={
	\node[PBS] (-x) at (0,0) {};
    \pic[rotate=-45,scale={0.6}] (-y1) at (0.5,0.5) {PNRD};
    \pic[rotate=-135,scale={0.6}] (-y2) at (0.5,-0.5) {PNRD};
    \draw[red] (-y1-c.center) -- (-x.center);
    \draw[red] (-y2-c.center) -- (-x.center);
    }
}

\node[BS] (BS1) at (2,1.5) {};
\node[BS] (BS2) at (3.5,3) {};
\node[PBS] (PBS1) at (2,4.5) {};
\node[PS] at (1.5,5) {};
\node[PS] at (2.5,4) {};

\foreach \x in {(5,4.5),(5,1.5),(3.5,0)} {
	\pic at \x {PRD};
}
\draw[red] (1,0.5) -- (5,4.5) ;
\draw[red] (1,2.5) -- (3.5,0) ;
\draw[red,text=black] (1,3.5) node[below,rotate=-90] {$\ket{1}$} -- (PBS1.center);
\draw[red,text=black] (1,5.5) node[below,rotate=-90] {$\ket{1}$} -- (5,1.5);

\draw[decorate,decoration={brace,amplitude=5},xshift=-2] (1,0.5) -- (1,2.5)
    node[midway,xshift=-13,rotate=-90] {$\ket{\beta}$} ;

\end{tikzpicture}
\vspace*{-1.5\baselineskip}
\end{figure}

Interestingly, with just two extra photons, we find two schemes, $(1,1,\tfrac14,\tfrac14)$ and $(1,\tfrac34,\tfrac12,\tfrac14)$, achieving a discrimination probability of $\PrbSu=5/8=0.625$. 
The first can be easily described as half of the 4-photon 
Ewert--van Loock scheme \cite{Ewert20143Ancillae} mentioned above 
and is described in  Figure~\ref{fig:halfewvan}.
It is especially relevant experimentally, since it is the simplest scheme achieving a success rate above $1/2$.
It was independently obtained by Ewert and van Loock \cite{vanLoock2017ImplementationsRepeater}. 
By ``halving'' in the same way the Ewert--van Loock 12-photon scheme that uses $4+8$ single photons and achieves a probability of $25/32$, we find a similar ``intermediate'' scheme with 4+4=8 extra photons, achieving $ \PrbSu = 49/64$.
Unfortunately, even using the cluster, numerically testing this scheme ($n=12, k=8$) proved to be unfeasible.

We notice (as in~\cite{Smith2018ApproachingPhotons}) that using an odd number $k+1$ of single photons in the ancilla does not improve the discrimination probability over the case with $k$ photons.
This is in line with the analytically-derived behavior for polarization-preserving interferometers of Section~\ref{sec:photbnd}.

Very recently, Smith and Kaplan~\cite{Smith2018ApproachingPhotons} tackled a similar problem, 
numerically optimizing linear optical Bell measurements with single photons ancillæ.
Their measurement were allowed to be ambiguous, and the chosen figure of merit was the classical mutual
information between state preparation and measurement.
Remarkably, despite this difference, we find corresponding results for ancillæ up to five single-photons; with six photons, they find a slight improvement of their mutual information, but the corresponding measurement is ambiguous \cite{Smith2018Private}.
Even if with six photons (Table~\ref{tab:results}) we could not find any scheme beyond $\PrbSu=3/4$---we collected more than 10\,000 optimizations---the polarization-preserving bound allows for a scheme with $\PrbSu\leq 13/16$; an improvement over $3/4$ is therefore not excluded.

\subsection{GHZ and W states}
We also checked the possible use of multipartite entangled states and ancillæ.
A three-photon GHZ state ancilla,
\[ \ket{\mathrm{GHZ}_3} = \frac{1}{\sqrt{2}} \Big( \ket{000} + \ket{111} \Big), \]
does not seem to help with respect to a simple Bell pair, as we still attain $3/4$ discrimination probability as optimum.
So does a $\mathrm{GHZ}_4$ state, at the expense of more computational power;
we wrongly expected the latter to be useful, given its use (along with a Bell pair) in the second iteration of Grice's scheme~\cite{Grice2011ArbitrarilyElements}.
The analytical polarization-preserving bound predicts $\PrbSu\leq 1/2$ for all $\ket{\mathrm{GHZ}_k}$ when $k\geq 3$. However, for odd $k$, the rotation of the polarization of a single photon by an angle of $±\tfrac{π}{4}$ raises this bound to 3/4. 
This value can be achieved by a trivial network applying a simple $\frac{π}4$ rotation on $k-2$ spatial modes of the ancilla, which leaves the
remaining two photons in the $\ket{Φ^{±}}$ states, which can be used as described above%
\footnote{The phase of the Bell pair is determined by the parity of the measurement of the $k-2$ photons, 
  and its effect is simply exchanges the photon patterns for the detection of  $\ket{Φ^+}$ and $\ket{Φ⁻}$.}
to achieve a 3/4 success probability. 

Another interesting state to investigate is the three-photon W state,
\begin{align}
\ket{\text{W}_3} = \frac{1}{\sqrt{3}} \Big( \ket{100} + \ket{010} + \ket{001} \Big).
\end{align}
Like $\mathrm{GHZ}_3$, it is a genuinely 3-party entangled state but, unlike all other states studied above, it is not a graph state, not even a stabilizer state.
Its specific symmetry is likely the source of the interesting results we find (end of Table~\ref{tab:results}).
Having the same number of horizontally polarized photons in each term, this state is as useful as the vacuum for polarization-preserving interferometers, as showed in Section~\ref{sec:vacuumextra}.
However, the rotation of the polarization of two photons by $\tfrac{\pi}4$ gives the higher bound $\PrbSu≤2/3$,
and further manipulation (see below) raises this bound to $\PrbSu≤3/4$.
The best optimum we find numerically, when we use a network with no extra vacuum modes ($n=10$), is $\PrbSu = 5/9$, significantly lower than the $3/4$ achieved with a simpler two-photon Bell pair.
This optimum is extremely rare (once in more than 20\,000 optimizations), and we observe the figure of merit in this case to suffer heavily from the problem described in eq.~\eqref{eqn:fomrel} about the relationship between $f(U)$ and $\PrbSu(U)$.

However, in this case we could find a better scheme by manipulating the state ``by hand''.
By measuring the last two spatial modes we can apply a transformation such that the remaining modes can be, depending on the result of the measurement, either in the state $\frac{1}{\sqrt{2}}(\ket{2H,0} - \ket{0,2V})$ or $\ket{\Phi^+}$.
Applying to these modes the same unitary of the one-Bell-pair $\PrbSu=3/4$ Grice strategy gives a scheme for $\ket{W_3}$ with $\PrbSu=7/12$.
While our optimization program correctly identifies this scheme as a local optimum when put in as starting point, an added Gaussian noise of average magnitude well below the requested convergence accuracy is sufficient for the optimization to diverge from it.
This numerical fragility may be the reason why we could not find this optimum through the optimization.
Applying the analytical bound to such transformed ancilla gives us $\PrbSu \leq 3/4$.
Interestingly, adding at least two vacuum modes ($n\geq 12$) allows the program to reach the better discrimination probability of 0.5785508(2).
Still, this is slightly below the manually-found 7/12.

\section{\label{sec:Conclusion}Conclusion}

In this work we have investigated the optimal success probability of a linear optical Bell 
measurement assisted by different kinds of input ancillary states $\ket{\Upsilon}$.
In Section~\ref{sec:UppBPolInd}, we showed how to obtain an upper bound from the input photon polarization distribution in $\ket{\Upsilon}$, when the network is restricted to polarization-preserving interferometers;
we noticed that the bound is tight for some published schemes.
With the aim of exploring the parameter space of generic interferometers, we developed in Section~\ref{sec:prog} a linear optical network simulator, capable of evolving a generic input state through the network and computing the analytical expression of the probabilities of each detection event in the output.
We then conducted a numerical search for the optimal value of $\PrbSu$ in for fixed $\ket{\Upsilon}$, and we discussed how to reduce the overall computational cost by exploiting some symmetries of the problem at hand.
We presented the results of both analytical and estimated numerical bounds in Section~\ref{sec:results}, and we recall them in Table~\ref{tab:results}.

Through both the analytical study and the numerical optimization we find evidences (but no proofs) for the optimality of known small schemes.
Some of them seem achievable experimentally in the short term, as they require as ancilla either a small number of photons or an additional Bell pair.
While restricted to the polarization-preserving case, the photon-number based analytical upper bound, saturated by Grice's schemes, is evidence for their optimality if resources are measured in terms of the number of extra ancillary photons.
In this setting, we have also shown that employing many copies of a Bell pair leads to a different (and worse) scaling than using Grice's states, giving interesting insights into the reason why the big GHZ-like states that appear in the schemes of~\cite{Grice2011ArbitrarilyElements,Ewert20143Ancillae} are needed.
Of course, eq.~\eqref{eq:Psep} being only a bound,
more research is needed to investigate its tightness, and whether near unity success can indeed be achieved.

As pointed out in the paper, some interesting cases lie beyond the computational capabilities at our disposal.
While there is still room for improvement, \eg\ by further optimization of the code and/or by employing more CPU time, our numerical approach is at least as hard as computing permanents of $k\times k$ submatrices of a unitary matrix.
As proved by Valiant~\cite{Valiant1979ThePermanent} and more recently pointed out by Arkhipov and Aaronson \cite{Aaronson2013TheOptics} in the context of linear optics, this task pertains to the complexity class $\mathsf{\#P}$-hard and is not believed to be solvable in polynomial time on a classical computer.
However the symmetry of the Bell states and the unambiguity constraints, which enforce a structure on the matrix entries---by 
imposing many null probabilities---may 
enable significant speedups (even exponential ones), even if the overall scaling could stay exponential.
Recent works in \cite[Appendix B]{Tichy2011EntanglementParticles} and \cite[Appendix D]{Shchesnovich2013AsymptoticBosons} suggest optimized algorithms for computing the  permanent of matrices with repeated columns/rows; they may help to improve our computation.

We conclude by noting that our simulator might be useful in exploring the power of linear optics in solving other types of problems.
Due to the flexibility of Python and of the separation between symbolic computation and numerical optimization, the program only requires minor modifications in order to be adapted to new tasks.
As a matter of fact, it has already been used during discussions with Chabaud \textit{et al.\@} in order to gain insight on the effect of Hadamard networks, helping in the design of linear optical swap-test~\cite{Chabaud2018ProgrammableOptics}.

\begin{acknowledgments}
We warmly thank the Quantum Information team of the LIP6 for their hospitality, and especially 
Ulysse Chabaud for stimulating discussions.
We acknowledge the use of the computing center MésoLUM of the LUMAT research federation (FR LUMAT 2764). 
AO acknowledges financial support from ANR project ANR-16-CE39-0001
and the Erasmus+ Traineeship Programme of the European Union. 
\end{acknowledgments}

\interlinepenalty=10000 
\bibliography{Mendeley.bib,extra.bib} 
\onecolumngrid 

\end{document}